\documentstyle[12pt]{article}

\newcommand{\be}{\begin{equation}}
\newcommand{\ee}{\end{equation}}
\newcommand{\bea}{\begin{eqnarray}}
\newcommand{\eea}{\end{eqnarray}}
\newcommand{\beasn}{\begin{sneqnarray}}
\newcommand{\eeasn}{\end{sneqnarray}}
\newcommand{\bref}[1]{(\ref{#1})}
\newcommand{\eps}{\epsilon}
\newcommand{\veps}{\varepsilon}

\newcommand{\der}[2]{\frac{\partial #1}{\partial #2}}

\newcommand{\gh}[1]{{\cal #1}}
\newcommand{\agh}[1]{\bar{\cal #1}}

\makeatletter
\def\secteqno{\@addtoreset{equation}{section}%
\def\theequation{\thesection.\arabic{equation}}}
\def\endsecteqno{\def\theequation{\@ifundefined{chapter}%
{\arabic{equation}}{\thechapter.\arabic{equation}}}}
\def\restric#1#2{{\left. #1 \right|_{#2}}}
\def\dif{{\rm d}}
\def\deriv{\@ifnextchar[{\@deriv}{\@deriv[]}}
   \def\@deriv[#1]#2#3{\mathchoice%
{{\dif^{#1}#2\over\dif{#3}^{#1}}}{{\dif^{#1}#2/\dif{#3}^{#1}}}%
{{\dif^{#1}#2\over\dif{#3}^{#1}}}{{\dif^{#1}#2/\dif{#3}^{#1}}}}
\def\tr{{\rm tr}}
\makeatother

\secteqno

\title{\bf Field-Antifield formalism for Anomalous Gauge Theories}

\author{\sc Joaquim Gomis\thanks{Bitnet QUIM @ EBUBECM1}
        and Jordi Par\'{\i}s\thanks{Bitnet PARIS @ EBUBECM1}\\ \\
        \address{Departament d'Estructura i Constituents
               de la Mat\`eria\\
        Universitat de Barcelona\\
        Diagonal, 647\\
        E-08028 BARCELONA\\
        Catalonia, Spain}}

\date{}

\begin{document}
\maketitle
\thispagestyle{empty}

\begin{abstract}

An extension of the Field-Antifield formalism to treat
anomalous gauge theories with a closed, irreducible classical gauge
algebra is
proposed. Introducing extra degrees of freedom, we construct the gauge
transformations for these new fields, the Wess-Zumino term and the
corresponding measure.

\end{abstract}

\vfill 
\vbox{\raggedleft
 April 1992\par
 Universitat de Barcelona preprint UB-ECM-PF 92/10\par\null}

\clearpage

\section{Introduction}
\hspace{\parindent}%

During the last few years, there has been a great amount of works trying
to quantize anomalous gauge theories. One of the first steps in this
direction was done by
Polyakov \cite{Pol81} in the path integral quantization of the bosonic
string in the presence of the conformal anomaly. More recently,
Jackiw and Rajaraman
\cite{JR85}, have shown that the chiral Schwinger model, despite the
anomaly, is consistent, unitary and amenable of particle interpretation.
On the other hand, Faddeev and Shatashvili \cite{FS86}, for the
non-abelian case, proposed the introduction of extra
degrees of freedom which at the canonical level cancel the anomalies,
turning the second class Gauss law constraints into first class ones. In
path integral terms, the cancellation of the anomalies could be obtained
by adding a Wess-Zumino term for the extra degrees of freedom, which
comes to compensate the non-invariance of the fermionic measure.
Later on, some authors \cite{BSV86}, \cite{HT87} showed that the extra
degrees of freedom and the Wess-Zumino term appear naturally in the
path integral approach if, when aplying the Faddeev-Popov procedure
\cite{FP69}, the
integration over the elements of the gauge group and the regularisation
of the measure are carefully taken into account.

A uniform description of the anomalies in the context of the general
lagrangian quantization scheme of Batalin and Vilkovisky \cite{BV83}
has recently been given in \cite{TNP89}. In this paper,
it was showed that anomalies are related to a violation of the master
equation and it was stressed the
importance of the regularisation scheme introduced to give sense to some
ill-defined expressions. More concretely, anomalies appear in this
formalism whenever the so-called quantum master equation can not be
solved in a local way. A proposal to solve this problem in the
Field-Antifield formalism, for the case of the chiral QCD in two
dimensions, was considered in \cite{BM91}. There, introducing the
elements of the gauge group as new degrees of freedom,
a local solution for the quantum master equation was obtained.

In the present paper we propose an
extension of the Field-Antifield formalism to treat in a generic way
anomalous gauge theories with closed, irreducible classical gauge
algebra.
First of all, we extend the original configuration space introducing
extra degrees of freedom and construct an extension of the proper
solution which allow us to obtain the gauge transformations for the new
fields. After that, an explicit solution for the quantum master equation
is given in terms of the anomalies of the theory.
Finally, we complete the program studying the measure for the new
degrees of freedom.

We have organized this paper as follows. In sect.2 we start giving a
general setting of the standard Field-Antifield formalism and of the
anomalies in this framework. Since along our development
some technical
tools about quasigroups are needed, to be self-contained, in sect.3
we present a summary of the results given in \cite{B80}.
In sect.4, we describe the above extension when all
the gauge symmetries are anomalous and in sect.5 we adapt the formalism
to the case when only a subgroup of the whole gauge group is anomalous.
Sect.6 is devoted to present some explicit examples and, in sect.7, we
give some conclusions. We conclude with one appendix, where
we show that the election of a
particular parametrization of the gauge quasigroup do not affect our
formalism.

\section{The standard Field-Antifield formalism}
\hspace{\parindent}%

Let us consider a classical system described by an action
$S_0(\phi^i)$, $i=1,\ldots,n$, invariant under the gauge
transformations
$$
   \delta\phi^i=R^i_\alpha(\phi)\veps^\alpha.
$$
In the Field-Antifield formalism, the set of classical fields
$\{\phi^i\}$ is enlarged
to a new set of "quantum" variables $\{\Phi^A\}$, $A=1,\ldots,N$,
including the ghosts, ghosts for ghosts, etc. Afterwards, we associate
with each field $\Phi^A$ an antifield $\Phi^*_A$, with opposite
statistics to those of the fields $\Phi^A$. In this space of fields and
antifields, it is defined an odd symplectic structure called
antibracket \cite{BV83}, with opposite properties to those of the usual
graded Poisson bracket, as
\be
(X,Y)\equiv\frac{\partial_rX}{\partial\Phi^A}\frac{\partial_lY}
{\partial\Phi^*_A}-\frac{\partial_rX}{\partial\Phi^*_A}
\frac{\partial_lY}{\partial\Phi^A}.
\label{antibracket}
\ee

Then, a bosonic functional $W(\Phi,\Phi^*)$ with dimensions of action is
introduced satisfying the equation
\be
   \Delta\exp\left\{\frac i\hbar W(\Phi,\Phi^*)\right\}=0,
\label{delta exp}
\ee
where $\Delta$ is a second order differential operator given by
\be
   \Delta=\frac{\partial_r}{\partial\Phi^A}\frac{\partial_l}
   {\partial\Phi^*_A}.
\label{delta}
\ee
The above equation (\ref{delta exp}) can be written in an equivalent
form, taking into
account of the definition of the antibracket (\ref{antibracket}), as
\be
   \frac12 (W,W)=i\hbar \Delta W,
\label{q master eq}
\ee
which is the so-called {\it quantum master equation}. It is important to
note that equation (\ref{q master eq}), as it stands, is very formal.
Indeed, the operator $\Delta$ (\ref{delta}), acting on local functionals
is proportional to $\delta(0)$. Therefore, as the primary
purpose is to look for local solutions of equation (\ref{q master eq}),
a suitable regularisation scheme should be introduced to give sense to
this equation.

To solve (\ref{q master eq}), the bosonic function $W$
is expanded in powers of $\hbar$
$$
   W=S+\sum^\infty_{p=1}{\hbar}^p M_p,
$$
and plugging it into the quantum master equation, one obtain the
following relations
\bea
     (S,S)&=&0,
\label{master eq}\\
     (M_1,S)&=& i\Delta S,
\label{m1}\\
     (M_p,S)&=& i\Delta M_{p-1}-\frac12\sum^p_{q=1}
     (M_q,M_{p-q}),\quad\quad p\geq 2,
\label{ordre superior}
\eea
where (\ref{master eq}), the so-called (classical) master equation,
provides all the relations defining the gauge algebra of the theory
after imposing some boundary conditions on $S$ \cite{BV85} \cite{FH}.

The requirement of correctness of the classical limit imposes the
following boundary condition on $S$
$$
  \restric{S(\Phi,\Phi^*)}{\Phi^*=0}=S_0(\phi),
$$
where $S_0(\phi)$ is the classical action of the theory.
On the other hand, as a consequence of (\ref{master eq}), one can show
that $S$ admits $2N$ gauge transformations, being only $(2N-r)\geq N$ of
them independents. The nondegeneracy of the functional integral requires
then that
$S$ be proper, i.e., $r=N$. In this case, these $N$ gauge invariances
allow us to eliminate precisely the $N$ antifields $\Phi^*_A$,
introducing the gauge
fixing fermion $\Psi$ and restricting the theory to the surface $\Sigma$
defined by
\be
   \Sigma :\Phi^*_A-\frac{\delta \Psi(\Phi)}{\delta \Phi^A}=0.
\label{gauge fixing}
\ee
In these conditions, the restriction of $S$ to $\Sigma$, $S_{\Sigma}$,
represents the full action of the theory, while the remaining terms
$M_p$, $p\geq 1$, constitute quantum corrections to the naive
integration measure. It should be noted that the requirement of
nondegeneracy dictates also the minimal content of the set of quantum
fields $\{\Phi^A\}$ as well as suplementary conditions on $S$ and on the
gauge fixing fermion $\Psi$.

After all that, the functional integral which generates the correct
Feynman rules can be expressed as
\be
  Z_\Psi=\int\gh D \Phi\exp\{\frac{i}{\hbar} W_\Sigma(\Phi)\},
\label{funcional generador}
\ee
with $W_\Sigma(\Phi)$ given by
$$
   W_\Sigma(\Phi)=W\left(\Phi,\Phi^*=\frac{\delta\Psi}
   {\delta\Phi}\right).
$$

Now, let us analize in which conditions the functional integral
\bref{funcional generador} is independent of the gauge
fixing choice. To this end, define the gauge fixed BRST transformation
\be
  \delta F=\restric{(F,W)}{\Sigma}\Lambda,
\label{brst trans}
\ee
where $\Lambda$ is an infinitesimal anticommuting constant parameter.
Performing a BRST transformation on \bref{funcional generador} with
parameter $\Lambda=i/\hbar\delta\Psi$, i.e.,
$$
  \delta\Phi^A=\restric{(\Phi^A,W)}{\Sigma}\frac{i}{\hbar}
  \delta\Psi,
$$
where $\delta\Psi$ is an infinitesimal variation of the gauge fixing
fermion $\Psi$, one can express the variation of $Z_\Psi$ under a change
of $\Psi$ as
$$
  Z_{\Psi+\delta\Psi}=Z_\Psi+\int\gh D\Phi\exp\left\{\frac{i}{\hbar}
  W_\Sigma(\Phi)\right\}
  \left\{-\frac{i}{2\hbar}
   \restric{[2i\hbar\Delta W-(W,W)]}{\Sigma}
  \right\}\frac{i}{\hbar}\delta\Psi.
$$
Therefore, $Z_{\Psi}$ will be independent of the gauge fixing, at least
at formal level, iff $W$ verifies the quantum master equation
(\ref{q master eq}).

With this result we arrive at a fundamental question in this formalism,
namely, the existence of local solutions for (\ref{q master eq}). It is
well known that, while (\ref{master eq}) admits local proper solutions
under certain conditions \cite{Hen91} \cite{GP92}, the
existence of local solutions for $M_p$, $p\geq 1$, in general is not
guaranteed. More precisely, if the equation (\ref{q master eq}) can not
be solved with local terms $M_p$, $p\geq 1$, the theory has anomalies
and, as a consequence, the generating functional will depend on the
gauge choice \cite{TNP89}.
Therefore, from this point of view, anomalies are related to the
non-existence of local solutions of the quantum master equation
(\ref{q master eq}).

Since in this paper we will always be dealing with irreducible gauge
theories with closed algebras, let us analize more closely this case.
The proper solution of the master equation is given in this situation by
\footnote{From now on, in order to avoid cumbersome notations, we will
          assume $\eps(\phi^i)=0$.}
\be
    S=S_0(\phi)+\phi^*_i R^i_{\alpha}(\phi)\gh C^\alpha+
    \frac12 \gh C^*_{\alpha}T^{\alpha}_{\beta\gamma}(\phi)
    \gh C^{\gamma}\gh C^{\beta},
\label{proper sol}
\ee
where $R^i_{\alpha}$, $T^{\alpha}_{\beta\gamma}$ are the generators of
the gauge transformations and the first order structure functions of the
gauge algebra, verifying
\bea
    &&R^i_{\alpha,j}R^j_{\beta}-R^i_{\beta,j}R^j_{\alpha}=
    R^i_{\gamma}T^{\gamma}_{\alpha\beta},
\nonumber\\
    && T^{\mu}_{\alpha\gamma}T^{\gamma}_{\beta\delta}-
    T^{\mu}_{\alpha\beta,i}R^i_{\delta}+
    (\mbox{cyclic perm. of $\alpha$, $\beta$, $\delta$})=0.
\nonumber
\eea

Now, if we calculate $\Delta S$, which corresponds to the logarithm of
the jacobian of the infinitesimal BRST transformation
(\ref{brst trans}), we obtain
\be
    \Delta S=(R^i_{\alpha,i}+T^{\beta}_{\beta\alpha})\gh C^\alpha\equiv
    a_{\alpha}\gh C^{\alpha}.
\label{delta s}
\ee
As we have said before, for this expression to make sense, a suitable
regularisation scheme must be introduced. Once this is done, we have two
possibilities:
\begin{itemize}
\item $\Delta S=0$. In this case, the quantum master
equation is simply solved by taking $M_p=0$, $p\geq 1$, and no anomalies
occur.

\item $(\Delta S)_{\rm reg}=A_{\alpha}(\phi)\gh C^{\alpha}$. Then the
theory is potentially anomalous. In this case
anomalies will occur if eq. \bref{m1} can
not be solved with a local $M_1$ term.
\end{itemize}

With respect the terms $M_p$, $p\geq2$, since in this type of theories
$S$ is only linear in the antifields and $\Delta S$ does not depend on
them, $M_1$ can be chosen to be a function of the classical fields only,
$M_1=M_1(\phi)$, and eq.(\ref{ordre superior}) are immediately solved
by taking $M_p=0$ for $p\geq 2$. Therefore, no higher order terms in
$\hbar$ appear.

Now, let us analyze more closely the equation for the $M_1$ term. Since,
as we have just argued, $M_1$ can be chosen to be $M_1=M_1(\phi)$,
(\ref{m1}) can be written in the form
$$
   \frac{\partial M_1}{\partial\phi^i}R^i_{\alpha}(\phi)=
   i A_{\alpha}(\phi).
$$
The general solution for this equation, given by Batalin and
Vilkovisky in \cite{BV83}, reads
\be
   M_1^{(0)}(\phi)=M_{1,{\rm inv}}(\phi)+i\int^1_0\dif t
  \,A_{\alpha}(g)(D^{-1})^{\alpha\beta}(g)\psi_{\beta}(\phi),
\label{solucio m1}
\ee
where $\psi_{\beta}$ are arbitrary gauge conditions, i.e., the matrix
$D_{\alpha\beta}\equiv\left(\frac{\partial\psi_{\alpha}}{\partial\phi^i}
R^i_{\beta}\right)$ is invertible; $(D^{-1})^{\alpha\beta}$ is the
inverse
of $D_{\alpha\beta}$; $g^i(\phi,t)$ is the solution of the ordinary
differential equation with initial value problem
$$
   \deriv {g^i}{t}= R^i_{\alpha}(g)
   (D^{-1})^{\alpha\beta}(g)\psi_{\beta}(\phi),\quad\quad\quad
   \restric{g^i(\phi,t)}{t=1}=\phi^i,
$$
and $M_{1,{\rm inv}}(\phi)$ is an arbitrary gauge invariant functional.
Being $\psi_{\alpha}$ and $R^i_{\alpha}$ usually local functionals,
containing spacetime derivatives of the fields, the matrix
$D_{\alpha\beta}$ will be a differential operator with non-local invers.
Therefore, one can conclude that, except for some particular situations,
the solution (\ref{solucio m1}) for $M_1^{(0)}(\phi)$ will be non local
in general, i.e., the theory will be anomalous and some of the classical
symmetries will be lost at quantum level.

Then, according to the general ideas about the quantization of anomalous
gauge theories \cite{Pol81} \cite{FS86} \cite{Z83} \cite{WZ71},
 it would be useful to develop a formalism in
which, from the very beginning, some extra degrees of freedom, related
to the parameters describing the gauge group, were introduced, in such a
way that the formalism allowed us to construct a general expression for
the Wess-Zumino term and to have a consistent quantization of the
theory. Following these ideas, in the present work we propose
a general extension of the Field-Antifield formalism to treat in a
generic way anomalous gauge theories with irreducible, closed gauge
algebra, i.e., the gauge group of the theory is a quasigroup \cite{B80}.
Since our development heavily relies on
the properties of these quasigroups, to be self-contained, in the next
section we present a brief introduction to this topic.

\section{Introduction to Quasigroups}
\hspace{\parindent}%

The quasigroup structure is a generalization of the Lie group structure,
introduced by Batalin \cite{B80}. At infinitesimal level, a
quasigroup is realized by a set of generators acting as differential
operators on functions of some initial variables. These generators obey
Lie algebra commutation relations with the difference that the
structure coefficients in this case depend on the initial variables. At
the level of finite transformations, the main difference between a Lie
group and a quasigroup relies on the modification of the composition
law, which now depends not only on the parameters of the
transformations,
but also on the initial conditions or variables.
A possible realization of the quasigroup structure are, for instance,
the hypersurfaces of the first class constraints in the phase space.

To introduce the concept of quasigroup, it will be useful to have in
mind a manifold $\gh M$ with coordinates $\phi^i$, $i=1,\ldots,n$. In
these conditions, let us consider
a continuous transformation acting on the coordinates of $\gh M$ given
by
$$
   \bar\phi^i=F^i(\phi,\theta),
$$
with $\theta^\alpha$, $\alpha=1,\ldots,r$ a set of real parameters.
\footnote{Also in this case, for simplicity, we will restrict
          ourselves to the bosonic case, i.e.,
          $\eps(\phi^i)=\eps(\theta^\alpha)=0$.}

Assume now that the transformations $F^i(\phi,\theta)$ satisfy the
following properties:
\begin{enumerate}
\item
For $\theta^\alpha=0$, we have the identity transformation
$$
   \phi^i=F^i(\phi,0).
$$
\item
The composition law between two finite transformations reads
\be
   F^i(F(\phi,\theta),\theta')=F^i(\phi,\varphi(\theta,\theta';\phi)),
\label{comp law}
\ee
where $\varphi^\alpha(\theta,\theta';\phi)$ is the
composition law of the quasigroup.
\item
 Left and right units coincide
\be
   \varphi^\alpha(\theta,0;\phi)=\varphi^\alpha(0,\theta;\phi)=
   \theta^\alpha.
\label{units}
\ee
\item
A modified associativity law holds
\be
   \varphi^\alpha(\varphi(\theta,\theta';\phi),\theta'';\phi)=
\varphi^\alpha(\theta,\varphi(\theta',\theta'';F(\phi,\theta));\phi).
\label{ass law}
\ee
\item
There exists an inverse transformation given by
$$
   \phi^i=F^i(\bar\phi,\tilde\theta(\theta,\bar\phi)),
$$
with the invers $\tilde\theta^\alpha(\theta,\bar\phi)$ satisfying the
relations
$$
   \varphi^\alpha(\tilde\theta(\theta,\bar\phi),\theta;\bar\phi)=
   \varphi^\alpha(\theta,\tilde\theta(\theta,\bar\phi);\phi)=0.
$$
\end{enumerate}

The above conditions 1)-5) define the structure called quasigroup. From
them, we will obtain some relations describing the quasigroup at
infinitesimal level.

The generators of the infinitesimal transformations for the
variables $\phi^i$ are
$$
  R^i_\alpha(\phi)\equiv\restric{\frac{\partial F^i(\phi,\theta)}
  {\partial \theta^\alpha}}{\theta=0}.
$$
Antisymmetrizing the second derivatives of the modified composition law
(\ref{comp law}) with respect to $\theta^\alpha$, $\theta^{'\beta}$ in
$\theta=\theta'=0$, one obtains the algebra
\be
    R^i_{\alpha,j}R^j_{\beta}-R^i_{\beta,j}R^j_{\alpha}=
    R^i_{\gamma}T^{\gamma}_{\alpha\beta},
\label{gauge algebra a}
\ee
with the structure functions $T^\gamma_{\alpha\beta}(\phi)$ defined as
$$
  T^\gamma_{\alpha\beta}(\phi)=\restric{
  \left[\frac{\partial^2\varphi^\gamma(\theta,\theta';\phi)}
  {\partial\theta^\beta\partial\theta^{'\alpha}}-(\alpha,\beta)\right]}
  {\theta=\theta'=0}.
$$

{}From the associativity law (\ref{ass law}), one can get
the generalized Jacobi identity for the
structure functions $T^\gamma_{\alpha\beta}$
\be
    T^{\mu}_{\alpha\gamma}T^{\gamma}_{\beta\delta}-
    T^{\mu}_{\alpha\beta,i}R^i_{\delta}+
    (\mbox{cyclic perm. of ($\alpha$, $\beta$, $\delta$}))=0.
\label{jacobi a}
\ee

The generators of the quasigroup in the $\phi$ representation, defined
as
\be
   \Gamma_\alpha=R^i_\alpha(\phi)\frac{\partial}{\partial\phi^i},
\label{gen gamma}
\ee
verify, according to (\ref{gauge algebra a}), the commutation
relations of a quasialgebra
$$
   [\Gamma_\alpha,\Gamma_\beta]=-
   T^\gamma_{\alpha\beta}(\phi)\Gamma_\gamma.
$$

Let us introduce now the matrices $\mu$, $\tilde\mu$ defined by
\be
   \mu^\alpha_\beta(\theta,\phi)=\restric{
   \frac{\partial\varphi^\alpha(\theta,\theta';\phi)}
   {\partial\theta^{'\beta}}}{\theta'=0},\quad\quad
   \tilde \mu^\alpha_\beta(\theta,\phi)=\restric{
   \frac{\partial\varphi^\alpha(\theta',\theta;\phi)}
   {\partial\theta^{'\beta}}}{\theta'=0},
\label{mu}
\ee
while we denote their inverses as $\lambda$ and $\tilde\lambda$
$$
  \lambda^\alpha_\gamma\mu^\gamma_\beta=\delta^\alpha_\beta,
   \quad\quad\quad
  \tilde\lambda^\alpha_\gamma\tilde\mu^\gamma_\beta=\delta^\alpha_\beta.
$$
Note that these matrices will always be invertible, at least locally,
because the property
\be
  \restric{\lambda^\alpha_\beta}{\theta=0}=
  \restric{\tilde\lambda^\alpha_\beta}{\theta=0}=
  \restric{\mu^\alpha_\beta}{\theta=0}=
  \restric{\tilde\mu^\alpha_\beta}{\theta=0}=\delta^\alpha_\beta,
\label{invers 0}
\ee
holds by virtue of eqs.(\ref{units}), \bref{mu}.

With the aid of the associativity law (\ref{ass law})
one obtain an analog of the Lie equation
\be
   \frac{\partial F^i}{\partial\theta^\alpha}=
   R^i_\beta(F)\lambda^\beta_\alpha(\theta,\phi),
\label{f derivada}
\ee
and the useful transformation rule for the generators
\be
   \frac{\partial F^i}{\partial\phi^j}R^j_\beta(\phi)=
R^i_\gamma(F)\lambda^\gamma_\alpha\tilde\mu^\alpha_\beta(\theta,\phi).
\label{trans gen}
\ee

On the other hand, deriving the same equation \bref{ass law} with
respect to the parameters of the quasigroup, one obtain the analog of
the Lie equation for the composition functions $\varphi^\alpha$
\be
  \frac{\partial\varphi^\alpha(\theta,\theta';\phi)}
  {\partial\theta^{'\gamma}}
  =\mu^\alpha_\beta(\varphi(\theta,\theta';\phi),\phi)
    \lambda^\beta_\gamma(\theta',F(\phi,\theta)),
\label{ultima}
\ee
and the following commutation relations for the elements of the
matrices $\mu$, $\tilde\mu$
\bea
    \mu^\alpha_{\delta,\beta}\mu^\beta_\gamma-
    \mu^\alpha_{\gamma,\beta}\mu^\beta_\delta &=&
    \mu^\alpha_\beta T^\beta_{\delta\gamma}(\bar\phi),
\label{algebra mu}\\
   \tilde\mu^\beta_\gamma(D_\beta\tilde\mu^\alpha_\delta)-
   \tilde\mu^\beta_\delta(D_\beta\tilde\mu^\alpha_\gamma)&=&-
    \tilde\mu^\alpha_\beta T^\beta_{\delta\gamma}(\phi),
\label{algebra mu t}
\eea
with the operator $D_\beta$ defined as
$$
   D_\beta\equiv\left(\frac{\partial}{\partial\theta^\beta}-
   R^i_\alpha(\phi)\tilde\lambda^\alpha_\beta(\theta,\phi)
   \frac{\partial}{\partial\phi^i}\right).
$$
Notice that in the case of a Lie group, $\mu$ and $\tilde\mu$ are the
generators of the Lie group acting on itself, as we can deduce from the
above equations and ({\ref{mu}).

Similar commutation relations for $\lambda$, $\tilde\lambda$, involving
the structure functions $T^\gamma_{\alpha\beta}$ are also fulfilled
\bea
    &&\der{\lambda^\alpha_\gamma}{\theta^\beta}-
    \der{\lambda^\alpha_\beta}{\theta^\gamma}-
    T^\alpha_{\mu\nu}(\bar\phi)
    \lambda^\mu_\beta\lambda^\nu_\gamma=0,
\label{algebra lambda}\\
   &&D_\beta\tilde\lambda^\alpha_\gamma-
     D_\gamma\tilde\lambda^\alpha_\beta+
T^\alpha_{\mu\nu}(\phi)\tilde\lambda^\mu_\beta
\tilde\lambda^\nu_\gamma=0,
\nonumber
\eea
which are the analogs of the Maurer-Cartan equation for a Lie group.

Some other important relations involving the matrices $\lambda$,
$\tilde\lambda$, $\mu$, $\tilde\mu$ are
\be
   \tilde\mu^\beta_\gamma(D_\beta\mu^\alpha_\delta)-
   \mu^\beta_\delta\frac{\partial\tilde\mu^\alpha_\gamma}
   {\partial\theta^\beta}=0,
\label{mu mu t}
\ee
or equivalently
\be
   \lambda^\delta_\gamma(D_\beta\mu^\alpha_\delta)-
\tilde\lambda^\delta_\beta\frac{\partial\tilde\mu^\alpha_\delta}
   {\partial\theta^\gamma}=0.
\label{mu lambda t}
\ee

{}From (\ref{f derivada}) and (\ref{trans gen}) one can verify that the
finite transformations for the variables $\phi^i$ satisfy the following
invariance property
\be
  \left(-\tilde\mu^\beta_\alpha\frac{\partial\; }{\partial\theta^\beta}+
  R^i_\alpha\frac{\partial\;}{\partial\phi^i}\right) F^i(\phi,\theta)=
  -\tilde\mu^\beta_\alpha D_\beta F^i(\phi,\theta)=0,
\label{invariancia f}
\ee
which will turn out to be very important in the next
section. Indeed, as we will see, for the case of anomalous gauge
theories, the above equation suggests to introduce the parameters of the
gauge quasigroup as extra variables, transforming infinitesimally under
the action of the gauge group as
\be
  \delta\theta^\alpha=-\tilde\mu^\alpha_\beta(\theta,\phi) \veps^\beta.
\label{a}
\ee
It is not difficult to verify that (\ref{a}) is the infinitesimal
transformation law corresponding to the action of the group on itself,
in such a way that the elements transform as inverses, i.e.,
\be
\bar\theta^\alpha=\varphi^\alpha\left[\bar\veps(\veps,\bar\phi),\theta;
   \bar\phi\right].
\label{finite theta}
\ee
Indeed, taking the finite transformation law for the variables $\phi^i$,
$F^i(\phi,\veps)$, and \bref{finite theta}, and plugging them into
$F^i(\phi,\theta)$, one can check that
$$
  F^i(\bar\phi,\bar\theta)=
  F^i(F(\phi,\veps),\varphi(\tilde\veps,\theta;F(\phi,\veps)))
  =F^i(\phi,\theta),
$$
which is the finite version of the infinitesimal invariance property
(\ref{invariancia f}).

At this point, a natural question arises: given a set of generators
$R^i_\alpha(\phi)$ and of structure functions
$T^\alpha_{\beta\gamma}(\phi)$ verifying the relations
(\ref{gauge algebra a}) and (\ref{jacobi a}), is it possible to
reconstruct, at least locally, the quasigroup law of transformation
using the equations (\ref{f derivada}), \bref{ultima} and
(\ref{algebra lambda})? Batalin \cite{B80} proved that this is
indeed the case. More precisely, if one uses the canonical
parametrization, defined as the one verifying the condition
$\lambda^\alpha_\beta\theta^\beta=\theta^\alpha$,
we can write the finite transformation for the variables $\phi^i$ in a
formal way as
$$
  F^i(\phi,\theta)=
  \left(\exp\{\theta^\alpha \Gamma_\alpha\}\right)\phi^i,
$$
where $\Gamma_\alpha$ are the generators of the quasigroup in the
$\phi^i$ representation (\ref{gen gamma}). For a more exhaustive study
of the quasigroup structure, we refer the reader to the original
reference \cite{B80}.

\section{Extended Field-Antifield formalism for Anomalous Gauge
         quasigroups }
\hspace{\parindent}%

Let us consider a classical system described by an action
$S_0(\phi^i)$, $i=1,\ldots,n$, invariant under the gauge
transformations
\be
   \delta\phi^i=R^i_\alpha(\phi)\veps^\alpha,\quad\quad\quad
   \alpha=1,\ldots,r,
\label{trans fi}
\ee
where $\veps^\alpha(x)$ are arbitrary functions of the space-time
variables. We will restrict ourselves to irreducible theories with
closed gauge algebras, i.e., we will assume that the generators
$R^i_\alpha$ are independent and that the relations
(\ref{gauge algebra a}) and (\ref{jacobi a}) are verified in any
space-time point.
Therefore, the gauge structure results to be that of the quasigroup
defined in sect.3 and the minimal proper solution of the
master equation is given by (\ref{proper sol})
$$
    S=S_0(\phi)+\phi^*_i R^i_{\alpha}(\phi)\gh C^\alpha+
    \frac12 \gh C^*_{\alpha}T^{\alpha}_{\beta\gamma}(\phi)
    \gh C^{\gamma}\gh C^{\beta}.
$$

Now, let us calculate $\Delta S$ (\ref{delta s}) and suppose that, after
having
introduced a consistent regularization scheme, e.g. Pauli-Villars
\cite{PV} \cite{VIJ}, $(\Delta S)_{\rm reg}$ is
$$
   (\Delta S)_{\rm reg}=A_{\alpha}(\phi)\gh C^{\alpha}, \quad\quad
   \mbox{with}\quad\quad A_{\alpha}\neq 0,\quad
    \forall\alpha=1,\ldots,r.
$$
Assume also that there is no local $M_1(\phi)$ verifying the quantum
master equation at one loop, (\ref{m1}). Therefore, we are in the
situation where all the gauge symmetries are anomalous or, what is the
same, all the classical symmetries are lost at quantum level.
In this case, it is well known that some of the degrees of freedom which
can be dropped out at classical level using the gauge symmetries,
survive at quantum level and must be taken into account (e.g. the
Liouville field in 2-D gravity). To reproduce this fact, we will enlarge
the
classical configuration space maintaining the gauge group structure and
we will quantize the resulting theory using the Field-Antifield
formalism.

\subsection{Gauge transformations for the extra fields}
\hspace{\parindent}%

Introduce some new fields $\theta^{\alpha}$, $\alpha=1,\ldots,r$ (i.e.,
as many as anomalous gauge symmetries we have), transforming
infinitesimally under the action of the gauge group as
$$
  \delta\theta^\alpha=-R^\alpha_\beta(\theta,\phi) \veps^\beta,
$$
where $\veps^\beta(x)$ are the parameters of the gauge
quasigroup,
in such a way that
{\it the quasigroup structure continues to be given by the structure
functions} $T^\alpha_{\beta\gamma}(\phi)$. In
other words, we extend the classical configuration space in such a way
that the set of variables $(\phi^i,\theta^\alpha)$ constitutes a new
representation of the classical quasigroup structure. Introducing the
compact notation for the fields and generators
$$
   \phi^a=(\phi^i,\theta^\alpha),\quad\quad
   R^a_\alpha=(R^i_\alpha,-R^\beta_\alpha),
$$
this means that the new generators $R^a_\alpha$ must verify the same
algebra as the old ones, i.e.,
\be
   R^a_{\alpha,b}R^b_\beta- R^a_{\beta,b}R^b_\alpha=
  R^a_\gamma T^\gamma_{\alpha\beta}(\phi).
\label{relacio}
\ee

In order to construct this realization it will be useful to extend the
Field-Antifield formalism and
associate to these new fields their corresponding antifields
$\theta^*_\alpha$. Then, consider the following action in the
extended space of all the fields and antifields
\be
  \tilde S=S- \theta^*_\alpha R^\alpha_\beta\gh C^\beta,
\label{extended proper sol}
\ee
where $S$ is the original proper solution (\ref{proper sol}).

The requirement that the new set of variables provides a representation
of the same gauge quasigroup structure
is equivalent in this language to the condition
$(\tilde S,\tilde S)=0$. After imposing this result, the conditions on
the generators $R^\alpha_\beta$ are given by the terms proportional to
the antifields $\theta^*_\alpha$. Explicitely,
\be
    R^\sigma_\gamma \tilde D_\sigma R^\alpha_\beta-
    R^\sigma_\beta \tilde D_\sigma R^\alpha_\gamma=
    -R^\alpha_\mu T^\mu_{\beta\gamma}(\phi),
\label{nova}
\ee
where we have defined the operator $\tilde D_\sigma$ as
$$
   \tilde D_\sigma=\left(\frac{\partial}{\partial\theta^\sigma}-
   R^i_\gamma(R^{-1})^\gamma_\sigma
   \frac{\partial}{\partial\phi^i}\right).
$$
Note that equation \bref{nova} is nothing but the relation
(\ref{relacio}) for $a=\alpha$, as it should be.

These conditions on $R^\alpha_\beta(\theta,\phi)$ are exactly the same
as the commutation relations (\ref{algebra mu t}) verified by the
generators of the quasigroup acting on itself $\tilde\mu^\alpha_\beta$,
(\ref{mu}).
Therefore, we conclude that the transformations for the extra
fields can be taken as
\be
  \delta\theta^\alpha=-\tilde\mu^\alpha_\beta(\theta,\phi) \veps^\beta.
\label{trans theta}
\ee
This result is very natural and allow us to interpret the new variables
introduced as the parameters of the quasigroup,
transforming under the action of this group as "inverses", as it has
been suggested in the previous section.

Now, let us study the conditions to be verified for a function
$G(\phi, \theta)$ to be gauge invariant. From the
transformations (\ref{trans fi}) and (\ref{trans theta}), we have
$$
  -\tilde\mu^\beta_\alpha\frac{\partial G}{\partial\theta^\beta}+
  R^i_\alpha\frac{\partial G}{\partial\phi^i}=
  -\tilde\mu^\beta_\alpha D_\beta G(\phi,\theta)=0\Rightarrow
  D_\beta G(\phi,\theta)=0.
$$
But from (\ref{invariancia f}), we know that
$D_\beta F^i(\phi,\theta)=0$.
So, we conclude that the expression of the finite
gauge transformations for the classical fields $\phi^i$ with parameters
$\theta^\alpha$ are gauge invariants in this extended theory.

In summary, introducing the parameters of the gauge group as new fields,
we manage to construct a new realization of the quasigroup with some gauge
invariant quantities constructed from the finite gauge transformations
of the classical fields. However, it should be stressed that the
classical
physical content of the extended theory is different from the original
one, as we have extended the configuration space maintaining the same
number of gauge tranformations. Therefore, our construction is prepared
to mimic already at classical level the fact that some symmetries are
lost at quantum level and that, as a consequence, some degrees of
freedom will survive in the quantization process.

Thus, in order to complete the programme, we must look for a BRST
invariant quantum theory, independent of the gauge fixing, i.e.,
we must solve the quantum master equation (\ref{q master eq}) and
evaluate $M_1$ in the extended configuration space
$(\phi^i,\theta^\alpha)$.

\subsection{The $M_1$ or Wess-Zumino term}
\hspace{\parindent}%

Let us now study the consequences of the introduction of the new
variables at the level of the quantum master equation.
Consider the modified action $\tilde S$ (\ref{extended proper sol}),
which from now on we will call extended proper solution, and compute
$\Delta\tilde S$. We have
$$
   \Delta\tilde S= \Delta S-\tilde\mu^\beta_{\alpha,\beta}\gh C^\alpha=
   a_\alpha\gh C^\alpha -\tilde\mu^\beta_{\alpha,\beta}\gh C^\alpha.
$$
So, in general the introduction of new degrees of freedom modifies
$\Delta S$. Note also that, since at this level we have not a kinetic
term for the $\theta^\alpha$ fields in the expression of $\tilde S$, we
can only regularize the original part $(\Delta S)$.

Now, in the extended space of fields and antifields, we introduce a new
bosonic object $\tilde W$ given by
$$
   \tilde W=\tilde S+\hbar \tilde M_1,
$$
verifying the quantum master equation (\ref{q master eq}). The
relations (\ref{master eq}) and (\ref{m1}) read in this case
\bea
     (\tilde S,\tilde S)&=&0,
\nonumber\\
     (\tilde M_1,\tilde S)&=& i\Delta\tilde S.
\label{m1 tilde}
\eea

To solve the equation for $\tilde M_1$ \bref{m1 tilde} we
split it in two pieces, $\tilde M_1= M_1+N_1$, in such a way that
\bea
     (M_1,\tilde S)=(i\Delta S)_{\rm reg}&=&i A_\alpha\gh C^\alpha,
\label{m1 original}\\
     (N_1,\tilde S)&=&-i\tilde\mu^\beta_{\alpha,\beta}\gh C^\alpha.
\label{m1 theta}
\eea
The functional $M_1(\phi,\theta)$ is interpreted as the Wess-Zumino term
since its BRST variation
gives the anomaly. On the other hand, the $N_1$ term is introduced in
order to compensate the non-invariance of the measure of the extra
fields under BRST transformations. A BRST invariant measure
for the $\theta^\alpha$ fields will be then
$$
   \left[\gh D\theta^\alpha\exp\{i N_1\}\right].
$$

Now, let us concentrate on the Wess-Zumino term $M_1(\phi,\theta)$.
The equation (\ref{m1 original}) can be written as
$$
  -\tilde\mu^\beta_\alpha D_\beta M_1(\phi,\theta)=
  \left(-\tilde\mu^\beta_\alpha\frac{\partial }{\partial\theta^\beta}+
  R^i_\alpha\frac{\partial }{\partial\phi^i}\right) M_1(\phi,\theta)=
  i A_\alpha(\phi).
$$
Taking into account that there always exists a (in general non
local) functional $M^{(0)}_1(\phi)$ satisfying this equation, as we
have described in sect. 2, (\ref{solucio m1}), together with the gauge
invariance of the functions giving the finite gauge transformations of
the classical
fields, $F^i(\phi,\theta)$, a general solution for $M_1$ is
$$
    M_1(\phi,\theta)=M^{(0)}_1(\phi)+G(F(\phi,\theta)),
$$
with $G(\phi)$ an arbitrary functional of the classical fields.

Now, since we are interested in obtaining the
Wess-Zumino term, we should look for suplementary conditions restricting
the form of $G$. As it is well known, in the usual anomalous gauge
theories, the Wess-Zumino terms are charactherized for being 1-cocycles
of the corresponding gauge group \cite{Z83}.
Thus, we will also demand $M_1$ to be a
1-cocycle of the gauge quasigroup, i.e, to verify
\be
   M_1(F(\phi,\theta),\theta')-
   M_1(\phi,\varphi(\theta,\theta';\phi))+
   M_1(\phi,\theta)=0\,\quad(\mbox{mod}\; 2\pi).
\label{1 cocicle}
\ee
In fact, one can see that by choosing
$G(\phi)=-M^{(0)}_1(\phi)$, that is, taking $M_1$ as
\be
    M_1(\phi,\theta)=M^{(0)}_1(\phi)-M^{(0)}_1(F(\phi,\theta)),
\label{m extes 2}
\ee
this expression satisfies (\ref{1 cocicle}).
Therefore, (\ref{m extes 2}) will be the expression of the
Wess-Zumino term in this formalism.
It should also be noted that the fact that the Wess-Zumino term can be
obtained as the difference of a non-local functional (the effective
action) and its gauge transformed is a well-known result \cite{Z83}.

Let us further elaborate the expression (\ref{m extes 2}) of
$M_1(\phi,\theta)$. We can write it in an alternative way introducing
an integration over a real parameter $t\in[0,1]$, as
$$
   M_1(\phi,\theta)=\int_0^1\dif t
   \frac{\dif\;}{\dif t}\;M_1^{(0)}(F(\phi,\theta t)),
$$
where $F^i(\phi,\theta t)$ is a finite gauge transformation of the fields
$\phi^i$ with parameter $\theta^\alpha t$. Aplying the chain rule and
using the expression (\ref{f derivada}), we have
$$
   M_1(\phi,\theta)=\int_0^1
  \left(\frac{\partial M_1^{(0)}}{\partial\phi^i} R^i_\beta\
   (F(\phi,\theta t))\right)
  \lambda^\beta_\alpha(\theta t,\phi)\theta^\alpha \dif t,
$$
and taking into account the fact that the variation of $M_1^{(0)}(\phi)$
gives the anomalies, we can write
\be
   M_1(\phi,\theta)=-i\int_0^1 A_\beta(
   (F(\phi,\theta t))\lambda^\beta_\alpha(\theta t,\phi)\theta^\alpha
   \dif t,
\label{m amb t}
\ee
where $\lambda^\beta_\alpha(\theta,\phi)$ is the inverse of the matrix
$\mu^\alpha_\beta(\theta,\phi)$, (\ref{mu}).

Finally, if we are working in the canonical parametrization of the gauge
quasigroup, the above expression for $M_1$ simplifies to
\be
   M_1(\phi,\theta)=
   -i\int_0^1 A_\alpha((F(\phi,\theta t))\theta^\alpha\dif t.
\label{m extes 4}
\ee
Therefore, at the end we have an expression of the Wess-Zumino action
$M_1(\phi,\theta)$ in terms of the anomalies of the theory
$A_\alpha(\phi)$. Let us note, also, that the expression for
the Wess-Zumino term given above (\ref{m extes 4}), for theories with a
Lie group structure, coincide with the well-known expressions given in
the literature \cite{Z83}.

To conclude this section we would mention that the expressions for
$M_1(\phi,\theta)$ and $\tilde S$ have been constructed using a
given parametrization of the quasigroup.
In appendix A, we will show that the expressions of the
Wess-Zumino term and of the extended proper solution are form invariant
under changes of the parameters describing the gauge group.

\subsection{The measure for the $\theta^\alpha$ fields}
\hspace{\parindent}%

The next step in solving the equation (\ref{m1 tilde}) is to find an
expression for the term $N_1(\phi,\theta)$.
It should be stressed that all the results in this part
will be formal, as no explicit regularization for the measure of the
$\theta^\alpha$ fields is considered. Such regularization will
drastically depend on the concrete form of the Wess-Zumino term for the
extra fields.

Consider then the equation for $N_1(\phi,\theta)$, (\ref{m1 theta}),
which expresses the non-invariance of this term under BRST
transformations
$$
     (N_1,\tilde S)\equiv\delta_B N_1=
     -i\tilde\mu^\alpha_{\gamma,\alpha}\gh C^\gamma.
$$
The right hand side of this equation can be formally written as
$$
     -i\tilde\mu^\alpha_{\gamma,\alpha}\gh C^\gamma=
     [-i\tilde\mu^\alpha_{\gamma,\sigma}\;
      \lambda^\beta_\alpha\mu^\sigma_\beta]\gh C^\gamma,
$$
and using the relation (\ref{mu mu t}) we have
$$
     [-i\tilde\mu^\sigma_\gamma (D_\sigma\mu^\alpha_\beta)
      \lambda^\beta_\alpha]\gh C^\gamma=
     [i\mu^\alpha_\beta\; \tilde\mu^\sigma_\gamma (D_\sigma
      \lambda^\beta_\alpha)]\gh C^\gamma=
     -i\mu^\alpha_\beta\delta_B\lambda^\beta_\alpha=
      -i\delta_B(\tr\ln\lambda^\alpha_\beta).
$$
A particular solution for $N_1(\phi,\theta)$ is then
$$
   N_1(\phi,\theta)= -i\tr\ln\lambda^\alpha_\beta=
      -i\ln(\det\lambda^\alpha_\beta),
$$
which after exponentiation becomes
$$
  \exp\{iN_1(\phi,\theta)\}= \det\lambda^\alpha_\beta(\theta,\phi).
$$
Therefore, we conclude that a suitable BRST invariant measure for the
$\theta^\alpha$ fields will be
$$
    \gh D G_L(\theta,\phi)\equiv
    [\gh D\theta\det\lambda^\alpha_\beta(\theta,\phi)],
$$
which coincides with the so-called left-invariant measure for
quasigroups introduced by Batalin in \cite{B80}.

In summary, by introducing
as many new fields as anomalous symmetries there exist,
we have managed to describe an anomalous gauge theory in terms of a BRST
invariant one with extra degrees of freedom.
In particular, this means that all the classical symmetries are
maintained at quantum level and that we can use all of them to fix
the gauge invariance of our theory. In the next section, after
having adapted the above procedure for the case of an
anomalous subgroup, we will briefly describe how this gauge fixing
procedure goes on in general.

\section{Extended Field-Antifield formalism for Anomalous Gauge
         subgroups }
\hspace{\parindent}%

In the precedent section, we have presented an extended Field-Antifield
formalism for the case when all the gauge symmetries are anomalous.
However, this is a very restrictive case, as we know that, for some
theories, by a judicious choice of the regulator and the $M_1$ term, it
may be possible to arrange for certain symmetries to be anomaly free.
Therefore, a modification of the general formalism proposed until now
should be considered.

So let us consider the more general situation when only some part of the
whole gauge quasigroup is anomalous. To this end, assume that our gauge
transformations can be splitted in two parts as
$$
  \delta\phi^i=R^i_\alpha\veps^\alpha=R^i_A\veps^A+R^i_a\veps^a.
$$
Moreover, suppose that a consistent regularization scheme can be found
such that the $A$ part of the group has no potential anomalies. In other
words, assume that the functional $M_1^{(0)}(\phi)$ \bref{solucio m1}
has a local part $M_{\rm l}$, which can absorb, if any, the $A$ part of
the potential anomalies, i.e.,
$$
  (M_{\rm l}, S)=i A_A\gh C^A,
$$
while its non-local part $M_{\rm nl}$ verifies
$$
  (M_{\rm nl}, S)=i A_a\gh C^a.
$$
Therefore, we assume the $a$ part of the group to be anomalous and
$(\Delta S)_{\rm reg}$ to be given, up to local counterterms, by
$$
  (\Delta S)_{\rm reg}=A_a\gh C^a.
$$

Let us investigate for one moment the consequences of this assumption
for our theory. Assume that to regularize the above expression of
$\Delta S$ we use the Pauli-Villars scheme, as described in
\cite{TNP89} \cite{VIJ} in the Field-Antifield formalism. In this
scheme,
it is well known that anomalies are related to the non-invariance of the
Pauli-Villars action, $S_{PV}$, under the anomalous symmetries. So, if
$(\Delta S)_{\rm reg}=A_a\gh C^a$, this means that our regulator,
i.e., the Pauli-Villars action $S_{PV}$, is invariant under type
$A$ transformations, $\delta(\veps^A)\; S_{PV}=0$. Now, consider the
commutator of two type $A$ transformations acting on this action. We
will have in general
$$
  0=[\delta(\veps^A_1),\delta(\veps^A_2)]\; S_{PV}=
  \delta(\veps^A_{12})\;S_{PV}+ \delta(\veps^a_{12})\;S_{PV}=
  \delta(\veps^a_{12})\; S_{PV}.
$$
Therefore, being $S_{PV}$ not invariant under type $a$ transformations
by hipothesis, we conclude that $\veps^a_{12}=0$ and, as a consequence,
that type $A$ transformations close between themselves.

This general analysis seems not to give additional restrictions for the
$a$ part. Nevertheless, from now on we will assume that the $a$ part is
also a subgroup.
Thus, in terms of the gauge generators, we will have the following
relations
\bea
    R^i_{A,j}R^j_B- R^i_{B,j}R^j_A&=&R^i_D T^D_{AB},
\nonumber\\
    R^i_{a,j}R^j_b- R^i_{b,j}R^j_a&=&R^i_d T^d_{ab},
\nonumber\\
    R^i_{A,j}R^j_b- R^i_{b,j}R^j_A&=&R^i_d T^d_{Ab}+R^i_D T^D_{Ab},
\nonumber
\eea
with no restrictions on the structure functions $T^d_{Ab}, T^D_{Ab}$.
Now, following the general procedure presented in the last section, we
will introduce some extra fields $\theta^a$ in our formalism, as many
as anomalous symmetries we have, and quantize the resulting theory.

\subsection{The gauge transformations for the extra fields $\theta^a$}
\hspace{\parindent}

The infinitesimal transformation law of the new fields
$\theta^a$ under the action of the gauge quasigroup will be taken as
$$
  \delta\theta^a=-\tilde\mu^{'a}_\beta(\theta^a,\phi)
  \veps^\beta,
$$
in such a way that the new set of fields $(\phi^i,\theta^a)$
constitutes a
{\it new representation of the classical quasigroup structure}.
{}From them, the extended proper solution is
\be
   \tilde S=S-\theta^*_a\tilde\mu^{'a}_\beta\gh C^\beta.
\label{a propia}
\ee

The requeriment $(\tilde S,\tilde S)=0$ determines the conditions on the
generators to be verified in order to have the desired representation.
These conditions read now
\be
   \left(\tilde\mu^{'b}_\gamma\frac{\partial}{\partial\theta^b}
   -R^i_\gamma\frac{\partial}{\partial\phi^i}\right)
   \tilde\mu^{'a}_\beta-(\beta,\gamma)=
   -\tilde\mu^{'a}_\alpha T^\alpha_{\beta\gamma}(\phi).
\label{algebra mod}
\ee
The form of the above equations and the fact that no $\theta^A$ fields
are supposed to appear in the formalism would suggest that the
generators $-\tilde\mu^{'a}_\beta(\theta^a,\phi)$
could be obtained by simply taking the general
expression (\ref{trans theta}) for $\alpha=a$ in $\theta^A=0$, that is,
$$
  \delta\theta^a=-\tilde\mu^a_\beta(\theta^a,\theta^A=0,\phi)
  \veps^\beta.
$$
However, by taking the commutation relations (\ref{algebra mu t}) in
$\theta^A=0$, one can see that these operators in general will not
verify the original gauge algebra. Indeed, for $\alpha=a$ we have
$$
   \restric{\left
   [\left(\tilde\mu^b_\gamma\frac{\partial}{\partial\theta^b}+
   \tilde\mu^B_\gamma\frac{\partial}{\partial\theta^B}
   -R^i_\gamma\frac{\partial}{\partial\phi^i}\right)
   \tilde\mu^a_\beta\right]}{\theta^A=0}-(\beta,\gamma)=-
   \restric{\tilde\mu^a_\nu T^\nu_{\beta\gamma}}{\theta^A=0},
$$
and the term
$\tilde\mu^B_\gamma\frac{\partial}{\partial\theta^B} \tilde\mu^a_\beta$
destroies in general the expected algebra. Therefore, we must look for
the generators following another procedure.

To this end, it will be useful to consider, first of all, some extra
results for a quasigroup constituted by two subgroups when we put some
of the parameters to zero, as it happens in our case.
So consider the composition functions of the quasigroup and split it in
two sets
$$
  \varphi^\alpha(\theta,\theta';\phi)=\{
  \varphi^a(\theta,\theta';\phi);   \varphi^A(\theta,\theta';\phi)\},
$$
where the indices $a$, $A$ correspond to the two subgroups.
Due to the fact that the $a$ part is a subgroup, we have
$$
  \varphi^B(\theta^a,\theta^A=0,\theta^{'b},\theta^{'B}=0;\phi)=0.
$$
{}From this relation we obtain
$$
  \tilde\mu^B_a(\theta^b,\theta^B=0,\phi)=
  \restric{\frac{\partial
  \varphi^B(\theta^{'a},\theta^{'A}=0,\theta^b,\theta^B=0;\phi)}
  {\partial\theta^{'a}}}{\theta^{'a}=0}=0,
$$
and proceeding in the same way one can see that
$\mu^B_a(\theta^b,\theta^B=0,\phi)=0$. These results imply that
$$
  \restric{\tilde\mu^A_D\tilde\lambda^D_B}{\theta^A=0}=\delta^A_B,
\quad\quad\quad
  \restric{\tilde\lambda^A_a}{\theta^A=0}=0,
$$
and similars relations for $\lambda$ and $\mu$.

On the other hand consider the $a$ part of the group as a group by itself.
Its composition functions, say
$\tilde\varphi^a(\theta^a,\theta^{'b};\phi)$, can be obtained from the
composition functions of the complet quasigroup as
$$
  \tilde\varphi^a(\theta^a,\theta^{'b};\phi)=
  \varphi^a(\theta^a,\theta^A=0,\theta^{'b},\theta^{'B}=0;\phi).
$$
Therefore, we can see that the block $\tilde\mu^a_b$, corresponding to
the subgroup $a$ by itself, is
\bea
   && \tilde\mu^a_b(\theta^a,\phi)=
  \restric{\frac{\partial
  \tilde\varphi^a(\theta^{'a},\theta^b;\phi)}
  {\partial\theta^{'b}}}{\theta^{'b}=0}=
\nonumber\\
  &&\restric{\frac{\partial
  \varphi^a(\theta^{'a},\theta^{'A}=0,\theta^b,\theta^B=0;\phi)}
  {\partial\theta^{'b}}}{\theta^{'b}=0}=
   \tilde\mu^a_b(\theta^a,\theta^A=0,\phi),
\nonumber
\eea
i.e., it can be obtained from the block $\tilde\mu^a_b$ of the matrix
$\tilde\mu^\alpha_\beta$ evaluated in $\theta^A=0$. From this result we
obtain also
$$
  \restric{\tilde\mu^a_d\tilde\lambda^d_b}{\theta^A=0}=\delta^a_b,
  \quad\quad\quad
  \tilde\lambda^a_b\tilde\mu^b_D+\tilde\lambda^a_B\tilde\mu^B_D=0,
$$
and similar relations for the corresponding blocks of $\lambda$ and
$\mu$. It is important to note also that, since the complete matrices
$\lambda$, $\tilde\lambda$, $\mu$, $\tilde\mu$ verify the relation
\bref{invers 0}, this implies, in particular, that
$\tilde\lambda^A_B$, $\tilde\mu^A_B$, $\tilde\lambda^a_b$,
$\tilde\mu^a_b$ and the corresponding blocks of $\lambda$ and $\mu$ are
invertible in $\theta^A=0$.

Let us try now to obtain the generators for the $\theta^a$ fields. To
this end, consider the finite transformations of the fields $\phi^i$
under the anomalous part of the group
$$
\tilde F^i(\phi,\theta^a)\equiv F^i(\phi,\theta^a,\theta^A=0),
$$
and study their transformation properties under the action of the whole
gauge quasigroup. In order to do that, it is useful to consider the
invariance property of the complete finite gauge transformations
$F^i(\phi,\theta)$,
\bref{invariancia f}, and particularize it to $\theta^A=0$. Then, for
$\alpha=a$ we simply obtain
\be
    \restric{\left(\frac{\partial\tilde F^i}{\partial\phi^j} R^j_a-
    \frac{\partial\tilde F^i}
     {\partial\theta^b}\tilde\mu^b_a\right)}{\theta^A=0}=0,
\label{a part}
\ee
which reflects the invariance of the finite gauge transformations
$\tilde F^i(\phi,\theta^a)$ associated to the $a$ part of the group if
we consider this subgroup by itself. On the other hand, taking eq.
\bref{invariancia f} for $\alpha=A$, we have
$$
    \restric{\left(\frac{\partial\tilde F^i}{\partial\phi^j} R^j_A-
    \frac{\partial\tilde F^i}
     {\partial\theta^b}\tilde\mu^b_A-
    \frac{\partial F^i}
     {\partial\theta^B}\tilde\mu^B_A\right)}{\theta^A=0}=0.
$$
In order to eliminate the explicit dependence on the $\theta^A$
parameters in this expression, we can use the analog of the Lie
equations for the quasigroup \bref{f derivada} to write
$$
    \restric{\frac{\partial F^i}
     {\partial\theta^A}\tilde\mu^A_B}{\theta^A=0}=
     \restric{R^i_D(\tilde F)\;\lambda^D_A\;\tilde\mu^A_B}
     {\theta^A=0}+
     \restric{R^i_b(\tilde F)\;\lambda^b_A\;\tilde\mu^A_B}
     {\theta^A=0}.
$$
After that, the transformation rules for the generators $R^i_\alpha$,
\bref{trans gen}, adapted to this case, besides the invariance property
of $\tilde F^i$ \bref{a part} and the properties of the matrices
$\lambda$, $\mu$, $\tilde\lambda$, $\tilde\mu$ described above,
enables us to express $R^i_b(\tilde F)$ as
$$
  R^i_b(\tilde F)=\restric{
\frac{\partial F^i}{\partial\phi^j} R^j_\beta
  \;\tilde\lambda^\beta_\sigma\;\mu^\sigma_b}{\theta^A=0}=
  \restric{
\frac{\partial F^i}{\partial\phi^j} R^j_d
  \;\tilde\lambda^d_e\;\mu^e_b}{\theta^A=0}=
  \restric{\frac{\partial F^i}{\partial\theta^d}
  \mu^d_b}{\theta^A=0}.
$$
So at the end, we can write the
following transformation property for $\tilde F^i(\phi,\theta^a)$
\be
    \left(\frac{\partial\tilde F^i}{\partial\phi^j} R^j_A-
    \frac{\partial\tilde F^i}
     {\partial\theta^a}\tilde\mu^{'a}_A\right)=
    R^i_B(\tilde F) M^B_A,
\label{a1 part}
\ee
with the matrices $\tilde\mu^a_A$ and $M^B_A$ defined by
\be
   \tilde\mu^{'a}_A(\theta^a,\phi)= \restric{(\tilde\mu^a_A+
   \mu^a_b\;\lambda^b_D\;\tilde\mu^D_A)}{\theta^A=0},\quad\quad\quad
   M^B_A= \restric{\lambda^B_D\tilde\mu^D_A}{\theta^A=0}.
\label{solucio generadors}
\ee

Thus, from the above relations \bref{a part}, \bref{a1 part}, it seems
plausible to take as generators for the $\theta^a$ fields the matrices
which come together with $\partial/{\partial\theta^a}$. Therefore,
we propose as gauge generators for the extra fields
\be
\tilde\mu^{'a}_b(\theta^a,\phi)=\tilde\mu^a_b(\theta^a,\theta^A=0,\phi),
\quad\quad\quad
   \tilde\mu^{'a}_B(\theta^a,\phi)= \restric{(\tilde\mu^a_B+
   \mu^a_b\;\lambda^b_D\;\tilde\mu^D_B)}{\theta^A=0}.
\label{gen modificats}
\ee

To conclude, we must verify
if the above generators (\ref{gen modificats}) give a representation of
the original gauge algebra \bref{algebra mod}. The verification of the
relations (\ref{algebra mod}) for $(\gamma,\beta)=(d,b)$ can be obtained
directly from the original algebra (\ref{algebra mu t}) evaluated in
$\theta^A=0$. Indeed, in this case we have
$$
   \restric{\left[\left(\tilde\mu^b_d\frac{\partial}{\partial\theta^b}+
   \tilde\mu^B_d\frac{\partial}{\partial\theta^B}
   -R^i_d\frac{\partial}{\partial\phi^i}\right)
   \tilde\mu^a_b\right]}{\theta^A=0}-(\beta,\gamma)=
   \restric{-\tilde\mu^a_\nu T^\nu_{bd}}{\theta^A=0}=
   \restric{-\tilde\mu^a_e T^e_{bd}}{\theta^A=0},
$$
and since $\tilde\mu^{'b}_d=\restric{\tilde\mu^b_d}{\theta^A=0}$ and
$\restric{\tilde\mu^B_d}{\theta^A=0}=0$, the above expression is nothing
but (\ref{algebra mod}) for $(\gamma,\beta)=(d,b)$. With respect to the
remaining commutation relations,
by using the definition of these generators (\ref{gen modificats}), the
commutation relations of the original theory (\ref{algebra mu t})
evaluated in $\theta^A=0$, the relations given above betweeen
the different blocks of the matrices $\lambda$,
$\tilde\lambda$, $\mu$, $\tilde\mu$ in $\theta^A=0$, and the relations
(\ref{mu mu t}), (\ref{mu lambda t}) adapted to this case, we have
checked that, indeed, they are verified. We do not give explicit
calculations about this part because they do not give much insight to
the question.

Finally, let us stress that with such elections the
transformations properties for the combinations
$\tilde F^i(\phi,\theta^a)$ \bref{a part}, \bref{a1 part} could be simply
written as
\be
   \delta_{(a)}\tilde F^i=0,\quad\quad\quad
   \delta_{(A)}\tilde F^i= R^i_B(\tilde F) \left(M^B_A\veps^A\right),
\label{inv tilde f}
\ee
i.e., while $\tilde F^i$ are invariant under the $a$ part of the group,
they transform as the classical fields $\phi^i$ with respect the $A$
part of the group. These properties will be crucial in the derivation of
the form of the Wess-Zumino term.

\subsection{The Wess-Zumino term}
\hspace{\parindent}

Let us try now to solve the quantum master equation for this case.
To this end, consider the extended proper solution $\tilde S$
\bref{a propia} and evaluate $\Delta\tilde S$. We have
$$
   \Delta\tilde S=\Delta S-\tilde\mu^{'a}_{\beta,a}\gh C^\beta.
$$
Divide now the corresponding $M_1$ term in two
pieces, $\tilde M_1=M_1+N_1$, each of one verifying
\bea
     (M_1,\tilde S)=(i\Delta S)_{\rm reg}&=&i A_a\gh C^a,
\label{m1a}\\
     (N_1,\tilde S)&=&-i\tilde\mu^{'a}_{\beta,a}\gh C^\beta.
\label{n1a}
\eea

Let us concentrate now on the equation \bref{m1a} for the Wess-Zumino
term in this case. Since no ghosts $\gh C^A$ associated with the $A$
subgroup appear in this equation, \bref{m1a} implies, actually, two
groups of equations, that is,
\bea
   &&\left(-\tilde\mu^{'b}_a\frac{\partial}{\partial\theta^b}
   +R^i_a\frac{\partial}{\partial\phi^i}\right)
    M_1(\phi,\theta^a)= i A_a(\phi),
\label{eq m1a}\\
   &&\left(-\tilde\mu^{'b}_A\frac{\partial}{\partial\theta^b}
   +R^i_A\frac{\partial}{\partial\phi^i}\right)
    M_1(\phi,\theta^a)=0.
\label{eq m1A}
\eea
The general solution for (\ref{eq m1a}), taking for granted the
existence of a non-local functional $M_1^{(0)}(\phi)$
\bref{solucio m1} for $A_\alpha\equiv A_a$ and
the invariance property of $\tilde F^i(\phi,\theta^a)$
under the $a$ part of the group, is
$$
   M_1(\phi,\theta^a)=M_1^{(0)}(\phi)+\tilde G(\tilde F(\phi,\theta^a)),
$$
with $\tilde G(\phi)$ an arbitrary functional of the classical fields.
As in
the previous section, the requirement that $M_1$ be a 1-cocycle, in this
case of the $a$ part of the group, restricts $\tilde G$ to be
$\tilde G=-M_1^{(0)}$. Therefore, the expression we propose for the
Wess-Zumino term reads
\be
   M_1(\phi,\theta^a)=M_1^{(0)}(\phi)-M_1^{(0)}(\tilde F(\phi,\theta^a)).
\label{wza}
\ee

On the other hand, it is not difficult to verify that (\ref{wza}) is
invariant under the $A$ part of the group, i.e., it satisfies
eq (\ref{eq m1A}). Indeed, taking into account that $M_1^{(0)}$ verifies
(\ref{eq m1a}), (\ref{eq m1A}) by construction
\be
  \frac{\partial M^{(0)}_1(\phi)}{\partial\phi^i}R^i_A=0,\quad\quad\quad
  \frac{\partial M^{(0)}_1(\phi)}{\partial\phi^i}R^i_a=i A_a(\phi),
\label{invariancia A m1}
\ee
we can write (\ref{eq m1A}) as
$$
   \left(-\tilde\mu^{'b}_A\frac{\partial}{\partial\theta^b}
   +R^i_A\frac{\partial}{\partial\phi^i}\right)
    M_1(\phi,\theta^a)=
  \restric{-\frac{\partial M^{(0)}_1(\phi)}{\partial\phi^i}}
  {\phi=\tilde F}
    \left(\frac{\partial\tilde F^i}{\partial\phi^j} R^j_A-
    \frac{\partial\tilde F^i}
     {\partial\theta^a}\tilde\mu^{'a}_A\right),
$$
and using the transformation property of $\tilde F^i$ under the $A$ part
of the subgroup (\ref{a1 part}), we can write the above equation as
$$
  \left(-\frac{\partial M^{(0)}_1(\phi)}{\partial\phi^i}
    R^i_B(\tilde F)\right) M^B_A=0,
$$
which is zero as a consequence of the invariance of $M_1^{(0)}$ under
type $A$ transformations (\ref{invariancia A m1}).

To conclude the study of the Wess-Zumino term, let us try to write it in
terms of the anomalies of the theory, $A_a(\phi)$. To this end, first of
all, it should be noted that (\ref{wza}) can be written as
\bea
   M_1(\phi,\theta^a)&=&
   M^{(0)}_1(\phi)- M^{(0)}_1(\tilde F(\phi,\theta^a))
\nonumber\\
   &=&M^{(0)}_1(\phi)- M^{(0)}_1(F(\phi,\theta^a,\theta^A=0))=
    M_1(\phi,\theta^a,\theta^A=0)
\nonumber
\eea
where $M_1(\phi,\theta^a,\theta^A=0)$ is the Wess-Zumino term
(\ref{m extes 2}) obtained in the previous section, evaluated in
$\theta^A=0$. Therefore, recalling
the general expression of $M_1$ in terms of the anomalies $A_\alpha$
(\ref{m extes 4}) we can write
\be
   M_1(\phi,\theta^a)\equiv M_1(\phi,\theta^a,\theta^A=0)
   =-i\int_0^1 A_a((\tilde F(\phi,\theta t))
   \lambda^a_b(\theta^a t,0,\phi)\theta^b
   \dif t.
\label{m1 sub}
\ee

If we work with canonical parameters, we have, for $\theta^A=0$
\be
   \theta^a=\restric{\lambda^a_\beta\theta^\beta}{\theta^A=0}=
   \lambda^a_b(\theta^a,\theta^A=0,\phi)\theta^b,
\label{normal parameters}
\ee
where $\lambda^a_b(\theta^a,\theta^A=0,\phi)$ are exactly the matrix
$\lambda^a_b(\theta^a,\phi)$ which one would obtain in considering the
$a$ subgroup by itself. Therefore, the parameteres
verifying (\ref{normal parameters}) correspond precisely to the
canonical parametrization of the $a$ subgroup. Then, in this
parametrization (\ref{m1 sub}) reads
\be
   M_1(\phi,\theta^a)
   =-i\int_0^1 A_a((\tilde F(\phi,\theta t)) \theta^a \dif t.
\label{m 1 sub canonica}
\ee
This concludes the construction of the
Wess-Zumino term for the case of an anomalous subgroup.

\subsection{The measure for the $\theta^a$ fields}
\hspace{\parindent}

Let us concentrate now on the
equation (\ref{n1a}) for the $N_1$ term. In order to compensate the
second term in $\Delta\tilde S$, we must modify the measure in a
suitable way. To do it, one can look for a suitable determinant to be
incorporated to the naive measure, as we did in sect.4, or to look
for some Fujikawa variables \cite{Fuji} giving the correct measure, at
least at formal level. Now, we will choose the second procedure.

Consider then some combinations of the fields $\phi^i$ and $\theta^a$,
$G^a(\phi,\theta^a)$, such that
$\frac{\partial G^a}{\partial\theta^b}$ is invertible. To construct
these new variables, take some independent functions of the classical
fields $\phi^i$, $G^a(\phi)$, and substitute $\phi^i$ by their gauge
transformed with parameters $\theta^a$, i.e., define
$$
   G^a(\phi,\theta)=G^a(\tilde F(\phi,\theta)).
$$
Then, as a consequence of the invariance property of
$\tilde F^i(\phi,\theta^a)$ (\ref{inv tilde f}) we have
$$
  \delta_{(a)} G^a(\phi,\theta)=
     \restric{\frac{\partial G^a}{\partial\phi^i}}{\phi=\tilde F}
   \delta_{(a)}\tilde F^i=0.
$$

On the other hand, the $A$ transformations for these new variables read
$$
  \delta_{(A)} G^a(\phi,\theta)=
     \restric{\frac{\partial G^a}{\partial\phi^i}}{\phi=\tilde F}
      \delta_{(A)} \tilde F^i=
    \left[\left(\frac{\partial G^a}{\partial\phi^i} R^i_A
(\tilde F)\right)\restric{\lambda^A_D\tilde\mu^D_B}{\theta_A=0}\right]
    \veps^B,
$$
where we have used the $A$ transformations for $\tilde F(\phi,\theta^a)$
(\ref{inv tilde f}).

Impose now the measure for these fields to be invariant under
infinitessimal type $A$ transformations. The jacobian for the
transformation $G^{'a}=G^{a}+\delta_{(A)}G^a$ is
$$
   \gh D G^{'a}=\gh D G^a\left(1+\frac{\partial(\delta_{(A)}G^a)}
   {\partial G^a}\right),
$$
and the requirement of invariance under these infinitesimal
transformations reads
$$
   \left(\frac{\partial(\delta_{(A)}G^a)}{\partial G^a}\right)=
   \frac{\partial}{\partial G^a}
    \left[\left(\frac{\partial G^a}{\partial\phi^i} R^i_A
    (\tilde F)\right)
    \restric{\lambda^A_D\tilde\mu^D_B}{\theta_A=0}\right]\veps^B=0.
$$
Now, using the chain rule and the transformations rules for the
generators $R^i_\alpha$ (\ref{trans gen}) adapted to $\theta^A=0$, we
can write the above condition as
$$
   \left\{\left[M^a_{A,j}R^j_c(\tilde F)+M^a_B(\tilde F) M^B_{N,b}
   \mu^b_c(M^{-1})^N_A(\phi,\theta)\right](M^{-1})^c_a(\tilde F)
   \right\}M^A_D(\phi,\theta)=0,
$$
where we have defined
$$
   M^a_A\equiv
   \frac{\partial G^a}{\partial\phi^i} R^i_A,\quad\quad\quad
   M^a_b\equiv
   \frac{\partial G^a}{\partial\phi^i} R^i_b,
$$
and $M^A_B$ is the matrix defined in (\ref{solucio generadors}).
Finally, taking into account of the relations (\ref{algebra mu}) and
(\ref{mu mu t}), one can verify that
$$
   M^B_{N,b}\mu^b_c(M^{-1})^N_A(\phi,\theta)=T^B_{cA}(\tilde F),
$$
which allows us to write the above equation as
\be
  (M^{-1})^b_a\left[M^a_{A,j}R^j_b+ M^a_B T^B_{bA}\right](\phi)=0,
\label{equacio g}
\ee
where we have substituted $\tilde F$ by $\phi$, as it no produces any
modification in the form of the equation. Therefore, in order to have an
invariant measure, at least at formal level, we shall use as new
variables some functions of the fields $G^a(\phi)$, verifying
(\ref{equacio g}) and evaluated in $\tilde F(\phi,\theta^a)$.

To conclude this study, it could be interesting to express the measure
for the variables $G^a$ in terms of the original variables. We have
\bea
   \gh D G&=&\det\left(\frac{\partial G^a}{\partial\theta^b}\right)
   \gh D \theta=\det\left(
   \frac{\partial G^a}{\partial\phi^i} R^i_c(\tilde F)\lambda^c_b
   \right)\gh D\theta=
\nonumber\\
   &&[\det M^a_b(\tilde F)][\det\lambda^a_b]\gh D \theta=
   [\det M^a_b(\tilde F)]\gh D G_L(\theta,\phi).
\nonumber
\eea
So, besides the expected left invariant measure for the $a$ subgroup, a
new determinant arises in order to compensate the variation of the left
invariant measure $\gh D G_L$ under type $A$ transformations. Finally,
let us note that the exponentiation of these determinants produces the
$N_1$ term, whose BRST variation cancels the second part of
$\Delta \tilde S$, $-\tilde\mu^a_{\beta,a}\gh C^\beta$, as expected.

Finally, to conclude the construction, let us briefly comment about the
gauge fixing procedure in this extended formulation.

\subsection{The Gauge Fixing}
\hspace{\parindent}%

In the standard Field-Antifield formalism, to fix the gauge, one
introduce first of all the so-called auxiliar sector
of fields, constituted by the antighosts $\bar\gh C^\alpha$, the
Nakanishi-Lautrup fields, $B^\alpha$, and its corresponding antifields,
$(\bar\gh C^*_\alpha, B^*_\alpha)$, and modify the extended proper
solution (\ref{extended proper sol}) or \bref{a propia} as
\be
  \tilde S_{\rm n.m.}=\tilde S+\bar\gh C^*_\alpha B^\alpha.
\label{no minima}
\ee
Then, as explained in sect. 2, a gauge fixing fermion is introduced,
which for this type of theories is usually taken to be of the form
$$
   \Psi=\bar\gh C^\alpha\chi_\alpha.
$$
Here $\chi_\alpha$ are the gauge fixing conditions, which in
principle can depen on all the fields.
In order to make contact with the standard results let us
take $\chi_\alpha=\chi_\alpha(\phi)$. In these conditions, the gauge
fixing surface $\Sigma$ (\ref{gauge fixing}) is given by
$$
  \phi^*_i=\agh C^\alpha\frac{\partial \chi_\alpha}{\partial\phi^i},
  \quad\quad \theta^*_\alpha=0,\quad\quad(\theta^*_a=0),\quad\quad
  \agh C^*_\alpha=\chi_\alpha,
$$
while the non minimal extended proper solution (\ref{no minima})
restricted to this surface is
\be
   \tilde S_\Sigma(\Phi)= S_\Sigma(\Phi)
   =S_0(\phi)+\agh C^\alpha
   \left(\frac{\partial \chi_\alpha}{\partial\phi^i}R^i_\beta\right)
   \gh C^\beta+B^\alpha\chi_\alpha.
\label{sgf}
\ee
With respect the Wess-Zumino term, as no
dependence on the antifields appears on it, its expression on $\Sigma$
is not modified.
Therefore, the final form of the basic functional integral is
\be
   Z_\Psi=\int \gh D\phi\gh D\gh C \gh D\agh C \gh D B
   \gh D G(\theta,\phi)
   \exp\{\frac{i}{\hbar}[S_\Sigma(\Phi)+\hbar M_1(\phi,\theta)]\},
\label{z fin}
\ee
where $M_1(\phi,\theta)$ is the corresponding Wess-Zumino term and
$\gh D G$ the invariant measure for the extra fields $\theta^\alpha$ or
$\theta^a$.

We would like to comment about the dependence
of the Wess-Zumino term on the regularization and on the gauge-fixing
choice. For definiteness, let us assume that we have used the
Pauli-Villars regularization scheme \cite{TNP89}. In this procedure, the
operators to be used to regularize the expression $\Delta S$ are
directly related to the
kinetic terms appearing in the gauge-fixed proper solution $S_\Sigma$
\bref{sgf}.
Therefore, these operators and, as a consequence, the anomalies obtained
from them, will depend on the gauge fixing conditions we choose. In this
way, the regularisation procedure introduces a "hidden" dependence on
the gauge fixing choice through the Wess-Zumino term.
Finally, note that to be consistent, in building the generating functional
$Z_\Psi$ \bref{z fin}, it seems that one should consider the same gauge
fixing conditions
as the previously used to regularize the expression of $\Delta S$. This
concludes our presentation of this extended version of the
Field-Antifield formalism for anomalous gauge theories.

In the next section we will ilustrate our procedure using two
well-known examples: the chiral QCD in two
dimensions, where all the gauge symmetries are anomalous, and the
bosonic string, as an example where some of the symmetries can be kept
anomaly free.

\section{Examples}
\hspace{\parindent}%

\subsection{Chiral QCD in two dimensions}
\hspace{\parindent}

Consider first of all a typical example of anomalous gauge
theory: the chiral QCD in two dimensions.
The classical action for this system
$$
   S_0(A_\mu;\psi,\bar\psi)=\int\dif^2 x \left[
   -\frac14 \tr F^{\mu\nu}F_{\mu\nu}+i\bar\psi D\hspace{-2mm}\slash
   \frac{(1-\gamma_5)}{2}\psi
   \right],
$$
is invariant under the (nonabelian) gauge transformations
$$
   \psi'=g^{-1}\psi,\quad\quad
   \bar\psi'=\bar\psi g,\quad\quad
   A'_\mu=g^{-1}\partial_\mu g+ g^{-1}A_\mu g,
$$
where $g$ is an element of a compact Lie group $G$, $g\in G$. The
matrices $T^a$, giving a representation of the Lie algebra of the
group and over which the trace operation is defined, are assumed to
verify the usual relations
\be
  \tr(T^aT^b)=-\frac12\delta^{ab},\quad\quad
  [T^a,T^b]=f^{abc}T^c,\quad\quad
  (T^a)^+=-T^a.
\label{t a}
\ee

The minimal proper solution of the master equation is given in this case
by
$$
  S=S_0+\int\dif^2 x\left[
  A^*_\mu D^\mu\gh C+\psi^*\gh C\psi-(\bar\psi^*)^\dagger
  \gh C^\dagger(\bar\psi)^\dagger
  +1/2f^{abc}\gh C^*_a\gh C_b\gh C_c  \right],
$$
where $A_\mu^*$, $\psi^*$, $\bar\psi^*$ are the antifields associated to
the classical fields and
$\gh C=\gh C^aT^a$ is the Lie algebra valued ghost associated with
the gauge parameters. Now, if we formally calculate $\Delta S$
we obtain
$$
  \Delta S\propto\int\dif^2 x \left[\tr \gh C\;\delta(0)\right].
$$
Therefore, being this term
proportional to $\delta(0)$, some consistent regularisation scheme
should be considered. Using the Fujikawa regularisation method or a
Pauli-Villars regularisation, as described in \cite{TNP89},
one gets for $(\Delta S)_{\rm reg}$
$$
  (\Delta S)_{\rm reg}= \frac{i}{4\pi}\int\dif^2 x\tr\left\{
   \gh C[\veps^{\mu\nu}\partial_\mu A_\nu-\partial_\mu A^\mu\right]
  \equiv {\cal A}\cdot \gh C.
$$
{}From this expression we extract the following consistent anomaly
\be
    {\cal A}=\frac{i}{8\pi}
  \left[\partial_\mu A^\mu-\veps^{\mu\nu}\partial_\mu
   A_\nu\right],
\label{anomalia}
\ee
which can not be compensated by a local $M_1$ term. Therefore all
the gauge group is anomalous and we are in the conditions assumed in
sect.4.

Now, we introduce some new fields
$\theta^a(x)$, related with the canonical parameters describing the
gauge group. However, when dealing with semisimple Lie groups it is
useful to work directly with the elements of the gauge group, written as
$g(x)=\exp\{\theta^a(x)T^a\}$, rather than with the parameters
$\theta^a$. In terms of the
Field-Antifield formalism this simply corresponds to a canonical
transformation. Note also that the correct identification, in view of
our conventions for the transformation of the $\theta^a$ fields, is not
the usual one, but
$$
  g(x)=\exp\{-\theta^a(x) T^a\}\quad\quad\mbox{or}\quad\quad
  g^{-1}(x)=\exp\{\theta^a(x) T^a\}.
$$
In this case, the finite gauge transformations in terms of $g$ read
$$
  g'(x)=g(x)h(x), \quad\quad\mbox{with}\quad\quad
  h(x)=\exp\{-\veps^a(x) T^a\}.
$$

The corresponding extended proper solution is
$$
  \tilde S= S+\int\dif^2 x \tr[g^*g\gh C],
$$
and no modification in $\Delta S$ appears. This is due to the
well-known fact that the naive measure for the fields $g$, ${\cal D}g$,
is given by
$$
  {\cal D}g\equiv {\cal D}\theta^a[\det\lambda^a_b(\theta)],
$$
that is, the one obtained in sect.4 adapted to the case of a semisimple
Lie group.

Finally, let us consider the general expression of the Wess-Zumino term
adapted to this case. If we define a smooth interpolation $g(t)$ between
$g(0)=1$ and $g(1)=g$ as
$$
   g^{-1}(t)=\exp\{t\theta^a T^a\}
$$
and take into account of the properties of the $T^a$ matrices
(\ref{t a}), we can write the Wess-Zumino term (\ref{m extes 4}) in
terms of Lie algebra valued quantitites as
$$
  M_1(\phi,\theta)= 2i\int_0^1\dif t \int\dif^2 x \tr\left\{
  \theta^a T^aT^b{\cal A}^b(F(\phi,g^{-1}(t))\right\},
$$
and due to the fact that $\theta^a T^a=g(t)\partial_t g^{-1}(t)$, we
arrive at
$$
  M_1(\phi,g^{-1})= 2i\int_0^1\dif t \int\dif^2 x
\tr\left\{g(t)\partial_tg^{-1}(t)\,
  {\cal A}(F(\phi,g^{-1}(t))\right\}.
$$
This is the well-known general expression of the Wess-Zumino term for
the case of semisimple Lie groups, in terms of the consistent
anomalies, given for example in \cite{Z83}.

Now, substituting in the above expression the consistent anomaly
${\cal A}$ (\ref{anomalia}), evaluated in the gauge transformed fields,
$$
   A'_\mu(t)=g(t)\partial_\mu g^{-1}(t)+ g(t)A_\mu g^{-1}(t),
$$
we arrive at
\be
  M_1(A_\mu,g^{-1})=\Gamma(g^{-1})-\frac1{4\pi}\int\dif^2 x
  \tr\left[g^{-1}\partial_\mu g \left(A^\mu-
  \veps^{\mu\nu}A_\nu\right)\right],
\label{mqcd}
\ee
where $\Gamma(g^{-1})$ is the well-known Wess-Zumino-Witten action
\cite{W84}, given by
\bea
   \Gamma(g^{-1})&=&
   -\frac1{8\pi}\int\dif^2 x \tr[\partial_\mu g^{-1}\partial^\mu g]
\nonumber\\
   &+&\frac1{4\pi}\int_0^1\dif t\int\dif^2 x
   \tr\left[g^{-1}\partial_t g g^{-1}\partial_\mu g g^{-1}\partial_\nu g
   \veps^{\mu\nu}\right].
\label{wzw}
\eea
Note that
if we work with the usual indices $+,-$, we can write the above
expression for $M_1$ \bref{mqcd} as
$$
  M_1(A_\mu,g^{-1})=\Gamma(g^{-1})-\frac1{4\pi}\int\dif^2 x
  \tr\left[g^{-1}\partial_- g A_+ \right],
$$
which coincides with the Wess-Zumino term given in
\cite{BM91} \cite{AR87}.

In conclusion, using our procedure we are able to construct a BRST
vacuumm functional, in such a way that all the gauge invariances of the
theory are maintained and they can be used to eliminate some degrees of
freedom. In this case, it is useful to introduce as a gauge condition
$$
     A_+=0.
$$
In this gauge, the fermions completely decouple from the
gauge field $A_\mu$, in such a way that its contribution, giving an
overall factor to the vacuum functional, can be dropped out. On the
other hand, the second term of the Wess-Zumino
term \bref{mqcd} vanishes. Therefore, at the end we arrive at the usual
formulation \cite{Pol2}, where the resulting effective theory is
described in terms of the Wess-Zumino-Witten action \bref{wzw} for the
elements of the gauge group.

\subsection{The Bosonic String}
\hspace{\parindent}%

Let us consider as a second example the bosonic
string, which falls in the class of systems described in sect.5. The
classical action for this system of $D$ bosons $X^\mu(\xi)$ coupled to
the gravitational field $g_{\alpha\beta}$ in two dimensions is
$$
   S_0=\int\dif^2\xi\left[-\frac12\sqrt{g}g^{\alpha\beta}
       \partial_\alpha X\partial_\beta X\right]\quad\quad
        \mbox{with}\quad\quad
  g\equiv-\det g_{\alpha\beta}.
$$
The gauge group of this system can be splitted in two parts, Weyl
transformations and diffeomorphisms, and the infinitesimal version of
these transformations reads
\bea
     \delta X&=&v^\alpha\partial_\alpha X,
\nonumber\\
     \delta g_{\alpha\beta}&=&\nabla_\alpha v_\beta+\nabla_\beta
     v_\alpha+\lambda g_{\alpha\beta},\quad\quad\quad
     \delta\sqrt{g}=\sqrt{g}(\nabla_\alpha v^\alpha+\lambda),
\nonumber
\eea
while the algebra of these infinitesimal transformations is
\bea
\nonumber
    [\delta_R(v^\alpha_1),\delta_R(v^\beta_2)]&=&
    \delta_R(v^\beta_2\partial_\beta v^\alpha_1-
    v^\beta_1\partial_\beta v^\alpha_2),
\\
\label{algebra corda}
    [\delta_R(v^\alpha),\delta_W(\lambda)]&=&
     \delta_W(-v^\alpha\partial_\alpha\lambda),\quad\quad
    [\delta_W,\delta_W]=0,
\eea
i.e., diffeomorphisms and Weyl transformations are subgroups of the
whole gauge group.

Consider now the minimal proper solution of the master equation. We have
\bea
    S= S_0+\int\dif^2 \xi&&\hspace{-5mm}
\left\{ X^*\gh C^\alpha\partial_\alpha X+
    g^{*\alpha\beta}\left(
    \nabla_\alpha \gh C_\beta+\nabla_\beta \gh C_\alpha+\gh C
     g_{\alpha\beta}\right)\right.
\nonumber\\
   &&\hspace{-5mm}
   \left.-\gh C^*_\beta\gh C^\alpha\partial_\alpha\gh C^\beta
    -\gh C^* \gh C^\alpha\partial_\alpha\gh C\right\},
\nonumber
\eea
where $\gh C^\alpha$, $\gh C$ are the ghosts associated with the
diffeomorphisms and Weyl tranformations, respectively, and $X^*$,
$g^{*\alpha\beta}$, $\gh C^*_\beta$, $\gh C^*$ are the corresponding
antifields.

Now, when we evaluate formally $\Delta S$, we obtain
$$
   \Delta S\propto\int\dif^2\xi\left[\partial_\alpha
   \gh C^\alpha\;\delta(0)
   \right],
$$
which should be
treated using a suitable regularisation scheme. But while in the other
example, the regulator did not preserve any gauge symmetry, here, by a
judicious choice of the regulator, it is possible to arrange for some
ghosts, Weyl or diffeomorphism ghosts, to be absent in the expression of
$(\Delta S)_{\rm reg}$.
Therefore, we are in the situation described in sect.5, namely,
the whole gauge group can be splitted in two subgroups, in such a way
that  one of them is anomalous free and the other is anomalous.
For definiteness, we will restrict
ourselves to the case when the Weyl subgroup is the anomalous one and
when the gauge fixing conditions correspond to the conformal gauge.

The regularized expression of $(\Delta S)_{\rm reg}$, which can be found
using a Fujikawa regularisation or a Pauli-Villars one preserving
diffeomorphisms, is given in this case by \cite{Fuj2}
\be
   (\Delta S)_{\rm reg}=\int\dif^2\xi
   \left[\left(\frac{D-2}{8\pi}\right)\gh M^2\sqrt{g}-
   \left(\frac{D-26}{48\pi}\right)\sqrt g R\right]\gh C\equiv\gh A
   \cdot\gh C
\label{anomalia conf}
\ee
where $R$ is the scalar curvature, $\gh M^2$ a mass regulator, $\gh C$
the Weyl ghost and $\gh A$ the conformal anomaly.

In this expression, the terms proportional to $D$, those coming
from the regularization of the $X$-loops, are gauge independent, as the
form of the kinetic term for the $X$ fields is independent of a gauge
fixing involving the metric field. However, in this system, the loops
involving the ghosts fields contribute also to this expression. In the
case we are considering, the conformal gauge, the form of the kinetic
term of the ghost system produces the $-26$ in front of the scalar
curvature. Had we choose another gauge fixing, for example
the (Polyakov) light-cone gauge, the contribution of the ghost-antighost
system would have been different. Therefore, this is an explicit example
which ilustrates the dependence of the anomalies on the gauge fixing due
to the regularization procedure.

Now, following the procedure we describe in sect.5, let us
introduce a new field $\theta$. Its gauge transformations read
\be
   \delta\theta=v^\alpha\partial_\alpha \theta-\lambda,
\label{trans liu}
\ee
and can be shown to verify (\ref{algebra corda}). From this form, we get
the extended proper solution
$$
  \tilde S=S-\int\dif^2\xi[\theta^*_\alpha
   (\gh C^\alpha\partial_\alpha \theta-\gh C)].
$$
Now, if we calculate $\Delta\tilde S$ we have two contributions:
$\Delta S$,
which will be compensated by the introduction of a suitable $M_1$ term,
and a contribution coming from the $\theta^*\theta$ part of $\tilde S$,
which indicates the non-invariance of the naive measure $\gh D \theta$
and the necessity of modifying it.

Let us construct, first of all, the $M_1$ term. Using the general
expression (\ref{m 1 sub canonica}) and the form of
the conformal anomaly (\ref{anomalia conf}), we have
$$
   M_1(g_{\alpha\beta},\theta)=-i\int_0^1\dif t
  \int\dif^2\xi
   \left[\left(\frac{D-2}{8\pi}\right)\gh M^2(\sqrt{g})'(t)-
   \left(\frac{D-26}{48\pi}\right)(\sqrt{g} R)'(t)\right]\theta,
$$
where $(\sqrt{g})'(t)$ and $(\sqrt{g}R)'(t)$ are the finite Weyl
transformations of these combinations of the metric field with parameter
$\theta t$, that is,
$$
   (\sqrt{g})'(t)=e^{\theta t}\sqrt{g},\quad\quad
   (\sqrt{g}R)'(t)=\sqrt{g}R+(\sqrt{g}\Box\theta)t,\quad\quad
   \Box\equiv g^{\alpha\beta}\nabla_\alpha\nabla_\beta.
$$
Plugging these transformations in the above expression of $M_1$ and
integrating over the real parameter $t$ we obtain
\bea
   M_1(g_{\alpha\beta},\theta)=-i\int\dif^2\xi\hspace{-5mm}
   &&\left\{\left(\frac{26-d}{48\pi}\right)
\left[\frac12\sqrt{g}\theta\Box\theta+\sqrt{g}R\theta\right]\right.+
\nonumber\\
   \hspace{-5mm}&&\left.\left(\frac{D-2}{8\pi}\right)
   \gh M^2(\sqrt{g}(e^\theta-1)\right\}.
\label{wz corda}
\eea

With respect the formal measure for the $\theta$ field, let us consider
a solution of the equation (\ref{equacio g}). One can easily see that
the function $G(g_{\alpha\beta})=g^{1/4}$ verifies this equation.
Therefore, a suitable Fujikawa variable to use in the measure will be
\be
   G(g'_{\alpha\beta})=(g^{1/4})'=
    (e^{\theta/2}g^{1/4})\equiv\varphi.
\label{variable fi}
\ee
This variable could be shown to be invariant under Weyl tranformations,
while the jacobian obtained after a reparametrization is formally $1$,
as expected.

In summary, we have constructed a theory which is anomaly free and BRST
invariant and, as a consequence, all the gauge invariances of the theory
can be gauge fixed. To be consistent with the regularisation procedure
described above, we should choose the "conformal gauge"
$$
  \chi_(g_{\alpha\beta})=[g_{\alpha\beta}-\tilde g_{\alpha\beta}],
$$
with $\tilde g_{\alpha\beta}$ an arbitrary background metric.

In this fixed background, (\ref{wz corda}) turns out to be the usual
Wess-Zumino term for the bosonic string in the conformal gauge
\cite{Pol81}, while
$\theta$ can be interpreted as the usual Liouville field. Note also that
the measure proposed for this field through the
variable $\varphi$ (\ref{variable fi}), corresponds exactly to the one
obtained using other procedures in several papers \cite{Fuj2}.

To conclude this construction, it would be interesting to obtain the
residual symmetry for the Liouville field $\theta$ from the original
gauge transformations (\ref{trans liu}). For simplicity, let us take
the flat background metric
$\tilde g_{\alpha\beta}=\delta_{\alpha\beta}$. As it is well known,
there exists some residual transformations which maintain
the gauge-fixing unaltered. This requirement imposes some conditions on
the parameters, which in this case read
$$
     \restric{\delta g_{\alpha\beta}}{\delta_{\alpha\beta}}=
     \partial_\alpha v_\beta+\partial_\beta
     v_\alpha+\lambda \delta_{\alpha\beta}=0.
$$
These conditions are verified if we take
$$
   \lambda=-\partial_\alpha v^\alpha\quad\quad\quad
   \mbox{and}\quad\quad\quad
     \partial_\alpha v_\beta+\partial_\beta
     v_\alpha=\delta_{\alpha\beta}(\partial_\sigma v^\sigma),
$$
which imply that $v^\alpha$ should be conformal killing vectors. Then,
the residual transformations for the $\theta$ field are
$$
  \delta\theta=v^\alpha\partial_\alpha\theta +\partial_\alpha v^\alpha,
$$
which are the usual transformations for the Liouville field.
This ends our formulation for the bosonic string.

\section{Conclusions}
\hspace{\parindent}%

An extension of the Field-Antifield formalism has been developed to
uncover the quantization of anomalous gauge theories with a classical
irreducible, closed gauge algebra. At the classical level, the main idea
is the construction of a new realization of the group structure in an
extended configuration space, constituted by the classical fields
$\phi^i$ and the parameters of the gauge group $\theta^\alpha$.

At the quantum level, we solve the quantum master equation with local
counterterms. More precisely, we obtain an explicit expression of the
Wess-Zumino term $M_1(\phi,\theta)$ in terms of the anomalies of the
theory using the finite gauge transformations, that coincides for
the case of a Lie group structure with the well-known results. This
Wess-Zumino term has in general a hidden dependence on the gauge-fixing
choice through the regularization procedure of $\Delta S$. Finally, a
formal expression for the measure of the extra variables has
been obtained.

The extension of this procedure to the case of open gauge algebras is a
subject under current study.

\appendix

\section*{Acknowledgements}
\hspace{\parindent}%

We are grateful to J.\,M.\,Pons and J.\,Roca for useful discussions.

This work has been partially supported by a NATO Collaborative
Research Grant (0763/87) and CICYT project no.\,AEN89-0347.

\section{Appendix A: Different parametrizations of the gauge quasigroup}
\hspace{\parindent}%

As it is well known, for the Lie groups there is no a unique
parametrization to describe its elements. Indeed, there exists
infinite sets of real parameters which describe equally well
the gauge group, and the same happens for a quasigroup. Therefore, a
natural question arises, namely, how can affect the election of a
particular parametrization to our formalism. As we will show,
to pass from one parametrization to another one correspond, in
the antibracket sense, to a canonical transformation in the extended
space of fields and antifields. But before showing this result, let us
consider some generalities concerning the change of parameters
describing the quasigroup, as described in \cite{B80}. For definiteness,
we will restrict ourselves to the case when all the gauge symmetries are
anomalous.

Suppose we are describing the quasigroup with a set of real parameters
$\psi^\alpha$, $\alpha=1,\ldots,r$. We ask ourselves how the relations
between the different functions defining the quasigroup are modified
when we pass to describe the quasigroup with another set of real
parameters $\theta^\alpha$ related to the old ones by some functions
$\psi^\alpha(\theta,\phi)$ verifying
\be
    \psi^\alpha(0,\phi)=0,\quad\quad\quad
    \det\left(\frac{\partial\psi^\alpha}{\partial\theta^\beta}\right)
    \neq0.
\label{conditions psi}
\ee
Following Batalin \cite{B80}, we define
$$
  \bar\phi^i=F^i(\phi,\psi(\theta,\phi))\equiv F^i_1(\theta,\phi),
$$
as the new transformations for the fields $\phi^i$. The
new composition functions $\varphi^\alpha_1(\theta,\theta';\phi)$ are
$$
  \varphi^\alpha(\psi(\theta,\phi),\psi(\theta',F_1(\phi,\theta);\phi))=
  \psi^\alpha(\varphi_1(\theta,\theta';\phi),\phi),
$$
while the new inverses $\tilde\theta_1^\alpha(\theta,\bar\phi)$ verify
the relations
$$
\phi^i=F^i_1(F_1(\phi,\theta),\tilde\theta_1^\alpha(\theta,\bar\phi)).
$$
The transformation law for the generators $R^i_\alpha$ is given by
\be
    R^i_{1\alpha}(\phi)\equiv
    \restric{\frac{\partial F^i_1(\phi,\theta)}{\partial\theta^\alpha}}
    {\theta=0}=
    \restric{\frac{\partial F^i}{\partial\theta^\beta}}{\psi=0}
    \restric{\frac{\partial\psi^\beta}{\partial\theta^\alpha}}{\theta=0}
    =R^i_\beta(\phi)\Omega^\beta_\alpha(\phi),
\label{gen omega}
\ee
where the matrix $\Omega^\beta_\alpha$ is defined as
$$
   \Omega^\beta_\alpha=
   \restric{\frac{\partial\psi^\beta}{\partial\theta^\alpha}}{\theta=0}.
$$

Batalin \cite{B80} shows that the new functions $F^i_1$,
$\varphi^\alpha_1$, $\tilde\theta^\alpha_1$, verify the same relations
as the old ones. Therefore, it seems that we can describe equally well
the quasigroup with the parameteres $\psi^\alpha$ or $\theta^\alpha$.
However, the generators of the gauge transformations for the classical
fields are different when using $\psi^\alpha$ or $\theta^\alpha$, as we
can see from (\ref{gen omega}). Because we suppose that the generators
are given from the very beginning, the relevant
parametrizations are the ones maintaining the form of the generators,
i.e., those verifying
\be
   \Omega^\alpha_\beta=\delta^\alpha_\beta.
\label{conditions psi 2}
\ee
So, from now on, we will only consider these special type of
parametrizations.

Now let us see how in the extended formalism this freedom in choosing
the parameters correspond to a canonical transformation. In the
parametrization $\psi^\alpha$, the extended proper solution is
\be
    \tilde S=S_0(\phi)+\phi^{'*}_i R^i_{\alpha}(\phi)\gh C^\alpha+
    \frac12\gh C^*_{\alpha}T^{\alpha}_{\beta\gamma}(\phi)
    \gh C^{\gamma}\gh C^{\beta}-
    \psi^*_\alpha\tilde\mu^\alpha_\beta(\psi,\phi)\gh C^\beta.
\label{solucio psi}
\ee
Change now the parametrization: $\psi^\alpha=\psi^\alpha(\theta,\phi)$.
In these conditions, the matrices
$\tilde\mu^\alpha_{1\beta}(\theta,\phi)$ and
$\tilde\mu^\alpha_\beta(\psi,\phi)$ corresponding to each
parametrization are related by \cite{B80}
$$
  \tilde\mu^\alpha_\gamma(\psi(\theta,\phi),\phi)=
  \tilde\mu^\beta_{1\gamma}(\theta,\phi)
  \left(\frac{\partial}{\partial\theta^\beta}-
   R^i_\sigma(\phi)\tilde\lambda^\sigma_{1\beta}(\theta,\phi)
   \frac{\partial}{\partial\phi^i}\right)\psi^\alpha(\theta,\phi).
$$

Substituting this relation in (\ref{solucio psi}), we can write the new
extended proper solution in exactly the same functional form
$$
    \tilde S=S_0(\phi)+\phi^*_i R^i_{\alpha}(\phi)\gh C^\alpha+
    \frac12\gh C^*_{\alpha}T^{\alpha}_{\beta\gamma}(\phi)
    \gh C^{\gamma}\gh C^{\beta}-
    \theta^*_\alpha\tilde\mu^\alpha_{1\beta}(\theta,\phi)\gh C^\beta,
$$
where the new antifields $\phi^*_i$, $\theta^*_\alpha$ are given by
\be
   \phi^*_i= \left(\phi^{'*}_i+\psi^*_\alpha
   \frac{\partial\psi^\alpha}{\partial\phi^i}\right),\quad\quad\quad
   \theta^*_\alpha=\psi^*_\beta
   \frac{\partial\psi^\beta}{\partial\theta^\alpha}.
\label{antifields}
\ee
It is not difficult to see by direct computation that the above
transformation for the antifields (\ref{antifields}), together with the
transformation for the fields
$$
  \phi^i=\phi^i,\quad\quad\quad
  \psi^\alpha=\psi^\alpha(\theta,\phi),
$$
is a canonical one in the antibracket sense \cite{BV83}\cite{VT82}.

Now, let us verify that
the Wess-Zumino term (\ref{m amb t}) is also form invariant under a
change of parameters. To this end, take this expression for $M_1$
and define $\theta^{'\alpha}=\theta^\alpha t$.
Then, we can write
\be
   M_1(\phi,\theta)=-i\int_0^\theta A_\beta
   ((F(\phi,\theta'))\lambda^\beta_\alpha(\theta',\phi)
   \dif\theta^{'\alpha}.
\label{m extes 3}
\ee
It should be
stressed that, although we have used a particular parametrization
($\theta'^\alpha=\theta^\alpha t$), it could be shown that the above
expression for $M_1$
(\ref{m extes 3}) does not depend upon the form of the integration path
over the gauge quasigroup manifold \cite{B80}.

Consider then the expressions for $M_1$ \bref{m extes 3}, with
parameters $\psi^\alpha$
$$
   M_1(\phi,\psi)=-i\int_0^\psi A_\beta
   ((F(\phi,\psi'))\lambda^\beta_\alpha(\psi',\phi)
   \dif\psi^{'\beta},
$$
and change the parametrization, $\psi^\alpha=\psi^\alpha(\theta,\phi)$.
In these conditions, the matrices
$\lambda^\alpha_\beta(\psi,\phi)$ and
$\lambda^\alpha_{1\beta}(\theta,\phi)$ corresponding to these
parametrizations are related by \cite{B80}
\be
   \lambda^\alpha_\beta(\psi,\phi)
   \frac{\partial\psi^\beta}{\partial\theta^\gamma}=
   \lambda^\alpha_{1\gamma}(\theta,\phi).
\label{lambda}
\ee
Taking into account of this relation and the fact that $\psi(0,\phi)=0$,
we conclude
$$
   M_1(\phi,\psi(\theta,\phi))\equiv
   M_1(\phi,\theta)=-i\int_0^\theta A_\beta
   ((F_1(\phi,\theta'))\lambda^\beta_\alpha(\theta',\phi)
   \dif\theta^{'\beta},
$$
i.e., the Wess-Zumino term has the same functional form in each
parametrization of the quasigroup verifying conditions
(\ref{conditions psi}) and (\ref{conditions psi 2}).

To conclude the proof, we should finally verify that the measure for the
extra fields $\theta^\alpha$ is also form invariant. Indeed, we have
\bea
     \gh D G_L(\psi,\phi)&=&[\gh D \psi
     \det\lambda^\alpha_\beta(\psi,\phi)]=
     \gh D\theta\;\det\left(\frac{\partial\psi^\alpha}
{\partial\theta^\beta}\right)\det\lambda^\alpha_\beta(\psi,\phi)
\nonumber\\
     &=&\gh D\theta \det\lambda^\alpha_{1\beta}(\theta,\phi)=
     \gh D G_L(\theta,\phi),
\nonumber
\eea
where use has been made of eq.\bref{lambda}.
This completes our proof that the election of a particular
parametrization of the quasigroup is unessential in our formalism.

\endsecteqno


\begin{thebibliography}{99}

\bibitem{Pol81}
{\sc A.\,M.\,Polyakov},
{\sl Phys.Lett.}\,{\bf B103} (1981), 207.

\bibitem{JR85}
{\sc R.\,Jackiw and R.\,Rajaraman},
{\sl Phys.Rev.Lett.}\,{\bf 54} (1985), 385.

\bibitem{FS86}
{\sc L.\,D.\,Faddeev}, {\sl Phys.Lett.}\,{\bf B145} (1984), 81;

{\sc L.\,D.\,Faddeev and S.\,L.\,Shatashvili},
{\sl Theor.Math.Phys.}\,{\bf 60} (1984), 206.

{\sc L.\,D.\,Faddeev and S.\,L.\,Shatashvili},
{\sl Phys.Lett.}\,{\bf B167} (1986), 225.

\bibitem{BSV86}
{\sc O.\,Babelon, F.\,A.\,Schaposnik and C.\,M.\,Viallet},
{\sl Phys.Lett.}\,{\bf B177} (1986), 385.

\bibitem{HT87}
{\sc K.\,Harada and I.\,Tsutsui},
{\sl Phys.Lett.}\,{\bf B183} (1987), 311.

\bibitem{FP69}
{\sc L.\,D.\,Faddeev and V.\,N.\,Popov},
{\sl Phys.Lett.}\,{\bf B25} (1967), 29.

\bibitem{BV83}
{\sc I.\,A.\,Batalin and G.\,A.\,Vilkovisky}, {\sl Phys.Lett.}
\,{\bf B102} (1981), 27; {\it Phys.Rev.}\,{\bf D28} (1983), 2567.

\bibitem{TNP89}
{\sc W.\,Troost, P.\,van Nieuwenhuizen and A.\,van Proeyen},
{\sl Nucl.Phys.}\,{\bf B333} (1990), 727.

\bibitem{BM91}
{\sc N.\,R.\,F.\,Braga and H.\,Montani}, "The Wess-Zumino term in the
Field-Antifield formalism", Rio de Janeiro University preprint,
IF-UFRJ-91-22, (1991).

\bibitem{B80}
{\sc I.\,A.\,Batalin}
{\sl J.\,Math.\,Phys.}\,{\bf 22} (1981), 1837.

\bibitem{BV85}
{\sc I.\,A.\,Batalin and G.\,A.\,Vilkovisky},
{\sl J.\,Math.\,Phys.}\,{\bf 26} (1985), 172.

\bibitem{FH}
{\sc J.\,M.\,L.\,Fisch and M.\,Henneaux},
{\sl Commun.Math.Phys.}\,{\bf 128} (1990), 627.

\bibitem{Hen91}
{\sc M.\,Henneaux},
{\sl Commun.Math.Phys.}\,{\bf 140} (1991), 1.

\bibitem{GP92}
{\sc J.\,Gomis and J.\,Par\'\i s}, "Perturbation Theory and Locality in
the  Field-Antifield formalism",
Uji Research Center preprint YITP/U-55, (1991).

\bibitem{Z83}
{\sc B.\,Zumino}, "Chiral anomalies and differential geometry", in
Relativity, groups and topology, eds. B.S.deWitt and R.Stora, Les
Houches session XL, North-Holland, Amsterdam (1983), and references
therein.

\bibitem{WZ71}
{\sc J.\,Wess and B.\,Zumino},
{\sl Phys.Lett.}\,{\bf B37} (1971), 95.

\bibitem{PV}
{\sc W.\,Pauli and F.\,Villars},
{\sl Rev.Mod.Phys.}\,{\bf 21} (1949), 434.

\bibitem{VIJ}
{\sc A.\,Diaz, W.\,Troost, P.\,van Nieuwenhuizen and A.\,van Proeyen},
{\sl Int.J.Mod.Phys.}\,{\bf A4} (1989), 3959.

\bibitem{Fuji}
{\sc K.\,Fujikawa},
{\sl Phys.Rev.Lett.}\,{\bf 42} (1979), 1195;
{\bf 44} (1980), 1733;
{\sl Phys.Rev.}\,{\bf D21} (1980), 2848.

\bibitem{W84}
{\sc E.\,Witten},
{\sl Commun.Math.Phys.}\,{\bf 92} (1984), 455.

\bibitem{AR87}
{\sc E.\,Abdalla and K.\,Rothe},
{\sl Phys.Rev.}\,{\bf D36} (1987), 3190.

\bibitem{Pol2}
{\sc A.\,Polyakov}, "Two dimensional quantum gravity", in Fields,
strings and critical phenomena, eds. E. Br\'ezin and J. Zinn-Justin, Les
Houches session XLIX, North-Holland, Amsterdam (1988).

\bibitem{Fuj2}
{\sc K.\,Fujikawa},
{\sl Phys.Rev.}\,{\bf D25} (1982), 2584.

\bibitem{VT82}
{\sc B.\,L.\,Voronov and I.\,V.\,Tyutin},
{\sl Theor.Math.Phys.}\,{\bf 50} (1982), 218.



\end{thebibliography}
\end{document}